\documentclass[oldversion]{aa}
\usepackage{txfonts}
\usepackage{graphicx}
\usepackage{natbib}
\begin{document}

\title{Pulsating hot O subdwarfs in $\omega$ Cen: mapping a unique instability strip on the Extreme Horizontal Branch \thanks{Based on observations collected at the European Organisation for Astronomical Research in the Southern Hemisphere, Chile (proposal IDs 083.D-0833, 386.D-0669, 087.D-0216 and 091.D-0791).}
}
\author{
S.K. Randall \inst{1}
\and A. Calamida \inst{2}
\and G. Fontaine \inst{3}
\and M. Monelli \inst{4}
\and G. Bono \inst{5}
\and M.L. Alonso \inst{6}
\and V. Van Grootel \inst{7}
\and P. Brassard \inst{3}
\and P. Chayer \inst{2}
\and M. Catelan \inst{6}
\and S. Littlefair \inst{8}
\and V.S. Dhillon \inst{8,4}
\and T.R. Marsh \inst{9}
}

\institute{ESO, Karl-Schwarzschild-Str. 2, 85748 Garching bei M\"unchen, Germany; \email{srandall@eso.org}
\and Space Telescope Science Institute, 3700 San Martin Drive, Baltimore, MD 21218, USA 
\and D\'epartement de Physique, Universit\'e de Montr\'eal, C.P. 6128, Succ. Centre-Ville, Montr\'eal, QC H3C 3J7, Canada
\and Instituto de Astrofisica de Canarias, Calle Via Lactea, 38205 La Laguna, Tenerife, Spain
\and Universit\`a di Roma ``Tor Vergata'', Department of Physics, via della Ricerca Scientifica 1, 00133 Rome, Italy
\and Pontificia Universidad Cat\'olica de Chile, Av. Vicu\~na Mackenna 4860, 782-0436 Macul, Santiago, Chile
\and Institut d'Astrophysique et de G\'eophysique de l'Universit\'e de Li\`ege, All\'ee du 6 Ao\^ ut 17, B-4000 Li\`ege, Belgium
\and Department of Physics and Astronomy, University of Sheffield, Sheffield S3 7RH, UK
\and Department of Physics, University of Warwick, Coventry CV4 7AL, UK
}
\date{Received date / Accepted date}

\abstract
{We present the results of an extensive survey for rapid pulsators among Extreme Horizontal Branch (EHB) stars in $\omega$ Cen. The observations performed consist of nearly 100 hours of time-series photometry for several off-centre fields of the cluster, as well as low-resolution spectroscopy for a partially overlapping sample. We obtained photometry for some 300 EHB stars, for around half of which we are able to recover light curves of sufficient quality to either detect or place meaningful non-detection limits for rapid pulsations. Based on the spectroscopy, we derive reliable values of $\log{g}$, $T_{\rm eff}$ and $\log{N(\rm He)/N(\rm H)}$ for 38 targets, as well as good estimates of the effective temperature for another nine targets, whose spectra are slightly polluted by a close neighbour in the image. 

The survey uncovered a total of five rapid variables with multi-periodic oscillations between 85 and 125 s. Spectroscopically, they form a homogeneous group of hydrogen-rich subdwarf O stars clustered between 48,000 and 54,000 K. For each of the variables we are able to measure between two and three significant pulsations believed to constitute independent harmonic oscillations. However, the interpretation of the Fourier spectra is not straightforward due to significant fine structure attributed to strong amplitude variations. In addition to the rapid variables, we found an EHB star with an apparently periodic luminosity variation of $\sim$2700 s, which we tentatively suggest may be caused by ellipsoidal variations in a close binary.

Using the overlapping photometry and spectroscopy sample we are able to map an empirical $\omega$ Cen instability strip in $\log{g}-T_{\rm eff}$ space. This can be directly compared to the pulsation driving predicted from the Montr\'eal "second-generation" models regularly used to interpret the pulsations in hot B subdwarfs. Extending the parameter range of these models to higher temperatures, we find that the region where $p$-mode excitation occurs is in fact bifurcated, and the well-known instability strip between 29,000--36,000 K where the rapid subdwarf B pulsators are found is complemented by a second one above 50,000 K in the models. While significant challenges remain at the quantitative level, we believe that the same $\kappa$-mechanism that drives the pulsations in hot B subdwarfs is also responsible for the excitation of the rapid oscillations observed in the $\omega$ Cen variables.
 
Intriguingly, the $\omega$ Cen variables appear to form a unique class. No direct counterparts have so far been found either in the Galactic field, nor in other globular clusters, despite dedicated searches. Conversely, our survey revealed no $\omega$ Cen representatives of the rapidly pulsating hot B subdwarfs found among the field population, though their presence cannot be excluded from the limited sample.
}

\titlerunning{EHB pulsators in $\omega$ Cen}
\authorrunning{S.K. Randall et al.}
\maketitle

\section{Introduction}

Hot subdwarfs are evolved compact stars with temperatures between $\sim$ 20,000--70,000 K and surface gravities in the 5.6 $\lesssim\log{g}\lesssim$ 6.1 range. They are commonly found both among the Galatic field population, where they are spectroscopically classified as sdB or sdO stars according to their temperature, and in globular clusters, where they are generally referred to as Extreme Horizontal Branch (EHB) or Blue Hook stars based on colour and magnitude measurements.  

There are several competing theories for the formation of hot subdwarfs. In the canonical scenario, the hot subdwarf progenitor loses nearly all of its envelope mass near the tip of the red giant branch (TRGB), leaving it with a H-envelope shell too thin to ascend the asymptotic giant branch after core He-ignition and producing a He-core burning star with a very thin hydrogen-dominated envelope \citep{dorman1993}. For stars that form part of a binary system, the necessary mass loss can be nicely modelled in terms of binary interactions involving Roche lobe overflow or a common envelope phase \citep{han2002,han2003}. Single hot subdwarfs may have undergone an unusually efficient stochastic mass loss, or can alternatively be explained in terms of merger scenarios involving at least one white dwarf \citep{han2008,justham2011,clausen2011}. For stars that lose sufficient mass before approaching the TRGB, the thermonuclear flash may be delayed until as late as the early white dwarf cooling track, leading to extensive mixing of thermonuclear burning products and leading to very hot subdwarfs with atmospheres dominated by He and showing enhancements of C and N \citep{d'cruz1996,brown2001,miller2008}. In globular clusters, an additional possibility is the He-enhanced scenario, where the high temperatures of EHB stars are attributed to a greatly enhanced initial He abundance in the globular cluster \citep[e.g.][]{d'antona2002}. 

The scenarios proposed can in principle be tested observationally, since the different formation channels leave an imprint on the resulting mass distribution, binary properties and atmospheric parameters. In this respect it is interesting that the hot subdwarf populations observed in different environments (Galactic field, globular cluster) appear to show systematic differences, both in terms of their distribution in $\log{g}$-$T_{\rm eff}$-$\log{N\rm(He)/N\rm(H)}$ space \citep[see e.g.][]{latour2014} and the observed binary fraction \citep{maxted2001,monibidin2008}. Another very promising way to observationally constrain the origin of hot subdwarfs is via the study of the pulsations detected in a subset of them. At the qualitative level, the presence or absence of pulsations can provide some indication of the internal structure of the star, while asteroseismology can be employed to measure specific stellar parameters such as the mass to a high precision \citep{fontaine2012}. 

Several types of non-radial pulsators are now known among hot subdwarfs in the Galactic field. The first to be discovered were the rapidly pulsating subdwarf B stars, also known as sdBV$_r$, V361 Hya, or EC 14026 stars \citep{kilkenny1997}. These objects are H-rich sdB stars found in a well-defined instability strip between $\sim$29,000 and 36,000 K and show multi-periodic luminosity variations on a typical timescale of 100--200 s with amplitudes of a few mmags (or a few tenths of a percent of the star's mean brightness). The non-adiabatic pulsation properties have been modelled very successfully in terms of low order pressure ($p$) modes excited by a $\kappa$-mechanism associated with a local overabundance of iron in the driving region. Indeed, the instability strip predicted by the so-called Montr\'eal ``second-generation'' models \citep{charp1996,charp1997} matches that observed almost perfectly. Interestingly, fewer than 10\% of sdB stars within the instability strip appear to show pulsations at a measurable level \citep{billeres2002,ostensen2010}. The most convincing explanation so far brought forward for this phenomenon is the perturbation of the levitating iron reservoir within the stellar envelope by weak winds \citep{fontaine2006}, but this remains to be modelled in detail. On the other hand, the quantitative interpretation of the observed period spectra based on adiabatic pulsation calculations is well-advanced for these stars and has led to the asteroseismic inference of fundamental parameters (including a very precise estimate of the mass, surface gravity, the thickness of the H envelope and in some cases the determination of the internal rotation) for some 15 targets \citep[see][for a recent review on sdB star asteroseismology]{charp2015}. 

The slowly pulsating subdwarf B stars (sdBV$_s$ or V1093 Her pulsators) were discovered some years after the sdBV$_r$ stars \citep{green2003}. Rather difficult to observe from the ground, these objects exhibit low-amplitude luminosity variations with periods on the order of one to two hours and are found at the cooler end of the sdB star distribution between $\sim$22,000 and 29,000 K. Like the sdBV$_r$ stars they have H-dominated atmospheres. The pulsations can be qualitatively explained in terms of low degree, high radial order gravity ($g$) modes excited by the same $\kappa$-mechanism that is active in the rapid pulsators \citep{fontaine2003}, however accurately reproducing the observed instability strip remains a challenge. Models incorporating not only iron but also other radiatively levitating elements such as nickel \citep{jeffery2006,hu2011,bloemen2014} have been most successful in this respect. With the availability of long uninterrupted time-series photometry of extraordinary quality, quantitative asteroseismology has become feasible also for these pulsators, allowing the burning core to be probed via the deeply penetrating $g$ modes for the first time \citep{val2010a,val2010b,charp2011}.

In addition to the two well-established classes of pulsating hot subdwarf stars we know several hybrid pulsators that lie at the intersection of the sdBV$_r$ and sdBV$_s$ instability strips and show both $p$ and $g$-mode pulsations \citep[e.g.][]{schuh2006}, as well as two apparently unique objects. The first, LS IV$-$14$^{\circ}$116 \citep{ahmad2005}, is an intermediate He-rich sdB that falls in the middle of the sdBV$_r$ instability strip but shows sdBV$_s$-like $g$ mode pulsations. So far, there has been no entirely satisfactory explanation for the pulsations observed. Quite intriguingly, it was recently shown to be a halo object \citep{randall2015}, which may indicate a distinct evolutionary history compared to the garden variety sdB stars. The second unique object is V499 Ser (often referred to as SDSS J160043.6+074802.9 in the literature), a hot \citep[$T_{\rm eff}\sim$ 68,500 K according to][]{latour2011} He-rich sdO star that exhibits very rapid multi-periodic oscillations with periods between 60 and 120 s \citep{woudt2006}. Exploratory non-adiabatic computations suggest that the pulsations may be driven by the same $\kappa$-mechanism that is at work in the sdBV$_r$ and sdBV$_s$ stars \citep{fontaine2008}. 

All the pulsators described so far belong to the Galactic field population. Given that EHB stars are commonly found also in globular clusters, an obvious next step was to search for counterparts to the field pulsators there. Unfortunately, obtaining high quality light curves for those EHB stars is extremely challenging from an observational point of view due to the faintness of the targets and the crowding of the field. Ground-based observations are feasible only for a small number of nearby clusters besides being restricted to the outer parts of the cluster, and space-based observing time e.g. on HST is very competitive. We discovered the first rapid EHB pulsator candidate serendipitously based on two hours of time-series photometry for a field in $\omega$ Cen that had been selected as a backup target during an unrelated run with SUSI2 at the NTT operated by ESO on La Silla, Chile \citep{randall2009}. This initial detection triggered an extensive follow-up survey, which revealed the candidate pulsator to be the prototype of a hitherto unknown type of EHB pulsator. First results, including the announcement of four rapid pulsators, were published in a short Letter by \citet{randall2011}. Here, we present an in-depth analysis of the completed survey.

\section{The search for rapid EHB pulsators}

\begin{figure}[t]
\centering
\includegraphics[width=9.0cm,angle=0]{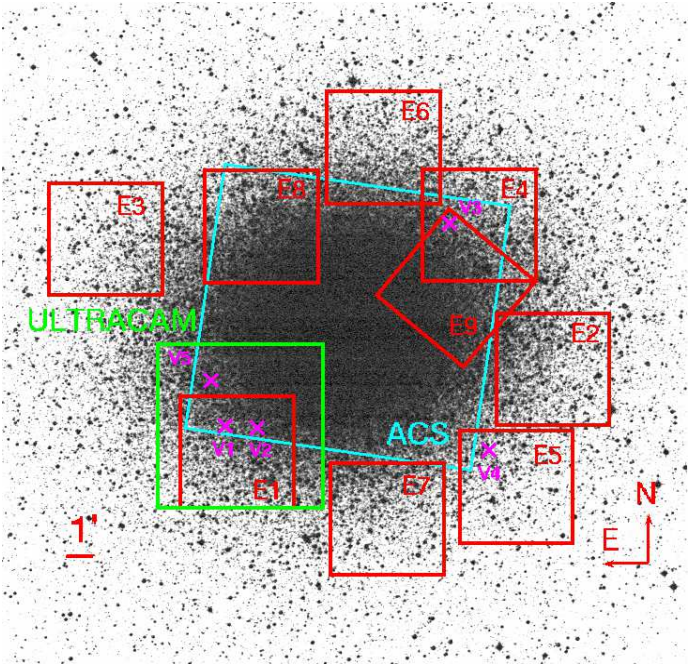}
\caption{Fields in $\omega$ Cen covered by the EFOSC2 (red) and ULTRACAM (green) observations. We also show the region covered by the ACS catalogue (turquoise) and indicate the variables detected from the survey (pink crosses).  
}
\label{fields}
\end{figure}

The time series photometry for $\omega$ Cen that we present here was obtained under different observing programs and spans three observing seasons in April 2009, 2011 and May 2013. All data were gathered at the NTT and were reduced and analysed in a very similar way. The main difference between the datasets is that the 2009 and 2013 runs were allocated on EFOSC2, while the 2011 data were obtained with ULTRACAM \citep{dhillon2007}. In total, we observed 10 pointings mapping part of $\omega$ Cen (see Fig. \ref{fields}) for $\sim$100 hours. Please see Table \ref{obslog} for an observing log containing the pointing coordinates as well as details on the time-series gathered. Details on the different observing runs are given in Sections 2.1 and 2.2.

The reduction procedure employed was as follows: all images were first bias-subtracted and flat-field corrected using standard IRAF routines (EFOSC2 data) and the ULTRACAM pipeline (ULTRACAM data). For the ULTRACAM data, a separate image was generated for each of the two amplifiers for the three ($u'g'r'$) chips. Then, a point-spread function (PSF) was calculated for every image using DAOPHOTIV \citep{stetson1987} on a selection of a few isolated and bright stars. The analytical part of the PSF was quadratically variable across the image and fixed to a Moffat function (EFOSC2) or dynamically chosen to be either a Moffat, a Gaussian or a Lorentz function by DAOPHOTIV (ULTRACAM). ALLSTAR was then run over all the images. In order to obtain a single photometric catalog for each observed field with the images in the same coordinate system, we used DAOMATCH/DAOMASTER \citep{stetson1994}. As a reference catalog, we employed UBVI-band photometry of $\omega$ Cen collected with the Wide Field Imager (WFI) on the 2.2-m ESO/MPI telescope at La Silla that covers an area of $\approx$ 40$\arcmin\times$ 42$\arcmin$ across the entire cluster \citep{castellani2007}. We then performed simultaneous PSF-photometry for each of the fields with ALLFRAME. For the ULTRACAM data the resulting catalogues for each observing night were scaled to a reference night (April 25), which corrects for nightly differences in the zeropoints due to the changing seeing conditions and the PSF. The EFOSC2 datasets were corrected only for the zeropoints relative to the WFI B-band. Note that no absolute flux calibration was performed since we are interested primarily in relative flux variations over time.

\begin{figure}[t]
\centering
\includegraphics[width=9.5cm,angle=0]{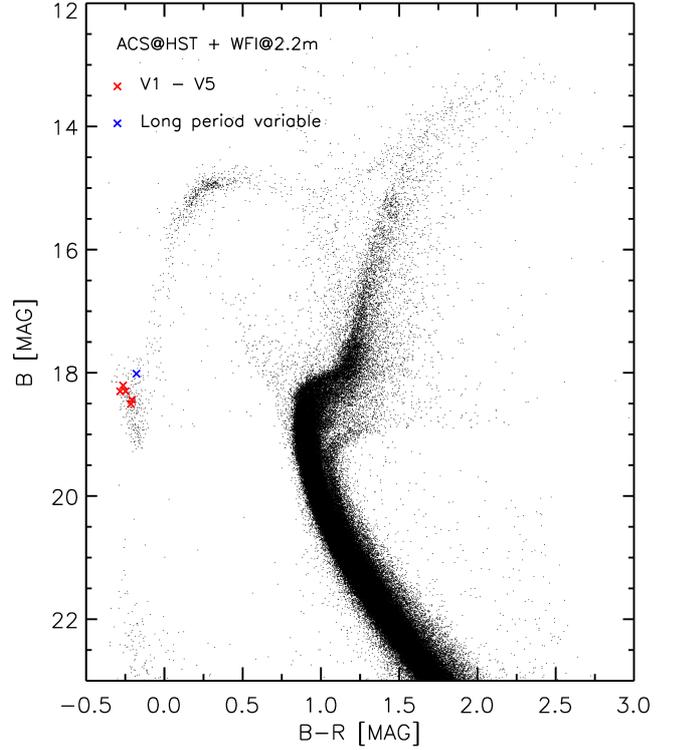}
\caption{Colour-magnitude diagram based on the merged ACS/WFI catalogue used for the selection of EHB star candidates in $\omega$ Cen. The position of the five short-period variables and the longer-period variable are indicated.
}
\label{cmd}
\end{figure} 

EHB star candidates were selected in brightness and colour from a merged catalogue comprising the WFI photometry mentioned above as well as F435W and F625W photometry obtained with the Advanced Camera for Surveys (ACS) aboard the Hubble Space Telescope \citep[for more details see][]{castellani2007}. The colour-magnitude diagram obtained on the basis of this merged catalogue is shown in Fig. \ref{cmd}. The ACS data have much better spatial resolution and photometric accuracy than the WFI measurements, but cover only the central part of the cluster (see Fig. \ref{fields}). Whenever our EFOSC2/ULTRACAM targets overlapped with the ACS catalogue, we used these measurements to select candidate EHB stars with brightness 17.8 $< F435W <$ 19.8 mag and colour $-$0.3 $ < F435W - F625W <$ 0.3 mag.  Otherwise we used the WFI catalogue and selected stars with 17.8 $< B <$ 19.8 mag and $-$0.3 $ < B-V < $ 0.2 mag. This process resulted in a list of $\sim$450 candidate EHB stars (see Table \ref{stats}) within the observed fields, 293 of which we were able to detect from our PSF photometry. Our EHB star sample should be understood to encompass bona fide core-helium burning EHB stars as well as post-EHB stars and a potentially mixed bag of post-RGB and post-AGB stars that happen to fall in the right region in the colour-magnitude diagram.

\begin{table*}[t]
\centering
\caption{Log of fast time-series photometry obtained for $\omega$ Cen}
\begin{tabular}{@{}ccccccc@{}}
\hline\hline
Field & $\alpha$ (2000.0) & $\delta$ (2000.0) & Start date (UT) & Start time (UT) & Length (h) & No. of images \\
\hline
ULTRACAM ($u'g'r'$) \\
\hline
ucam & 13:27:08.5 & $-$47:32:29 & 2011-04-22 & 23:16 & 9:47 & 5525 \\
ucam & 13:27:08.5 & $-$47:32:29 & 2011-04-23 & 23:16 & 10:14 & 5900 \\ 
ucam & 13:27:08.5 & $-$47:32:29 & 2011-04-24 & 23:19 & 9:48 & 6086 \\
ucam & 13:27:08.5 & $-$47:32:29 & 2011-04-25 & 23:53 & 9:12 & 5709 \\
ucam & 13:27:08.5 & $-$47:32:29 & 2011-04-27 & 23:08 & 9:47 & 6082 \\ 
\hline
EFOSC2 ($B$) \\
\hline
efosc1 (1) & 13:27:09.0 & $-$47:33:26.8 & 2009-04-15 & 23:58 & 6:03 & 597 \\
efosc1 (2) & 13:27:09.0 & $-$47:33:26.8 & 2009-04-17 & 02:05 & 5:05 & 481 \\
efosc2 & 13:26:01.7 & $-$47:30:06.3 & 2009-04-16 & 06:08 & 3:37 & 342 \\
efosc3 (1) & 13:27:38.3 & $-$47:25:46.3 & 2009-04-17 & 07:14 & 2:28 & 233 \\
efosc3 (2) & 13:27:38.3 & $-$47:25:46.3 & 2009-04-18 & 23:31 & 2:36 & 217 \\
efosc4 & 13:26:18.4 & $-$47:24:55.2 & 2009-04-19 & 02:12 & 4:10 & 385 \\
efosc5 & 13:26:08.9 & $-$47:34:26.7 & 2009-04-17 & 23:26 & 3:36 & 354 \\
efosc6 & 13:26:39.4 & $-$47:22:10.2 & 2009-04-18 & 03:37 & 2:49 & 244 \\
efosc7 & 13:26:36.4 & $-$47:35:44.7 & 2009-04-18 & 06:27 & 3:12 & 280 \\
efosc8 & 13:27:05.0 & $-$47:25:11.0 & 2009-04-19 & 06:24 & 3:20 & 339 \\
\hline
efosc9 (1) & 13:26:21.7 & $-$47:27:14.0 & 2013-05-11 & 01:34 & 4:55 & 426 \\
efosc9 (2) & 13:26:21.7 & $-$47:27:14.0 & 2013-05-11 & 23:11 & 2:55 & 249 \\
\hline
\end{tabular}
\label{obslog}
\end{table*}

\begin{figure}
\centering
\includegraphics[width=7.5cm,angle=90, bb = 40 80 550 600]{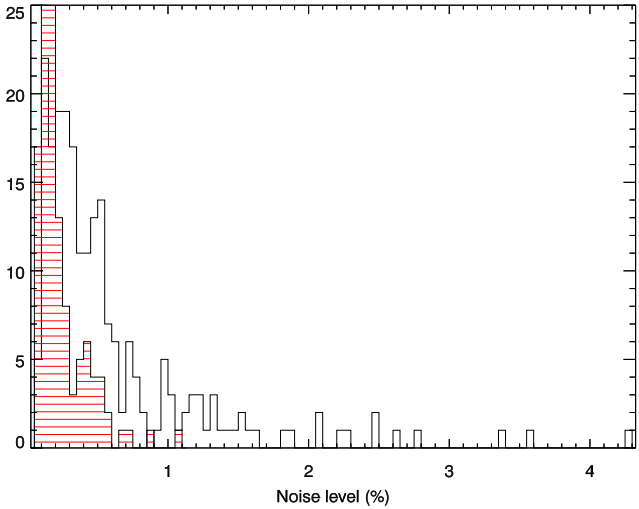}
\caption{Histogram of the noise level in the Fourier spectrum of the EHB lightcurves extracted from the EFOSC2 fields (non-filled area) and from the ULTRACAM field (red hashed area). The vast majority of the ULTRACAM targets have noise levels below 0.6 \% of the mean brightness of the star, while for the EFOSC2 fields there are several outliers where the quality of the light curve is low. Note that the noise level of the ULTRACAM data was measured in the $u'$ band, while for the EFOSC2 data it is based on the $B$ band.
}
\label{noise}
\end{figure}  

\subsection{A multi-field search for EHB pulsators with EFOSC2}

The 2009 observations obtained with EFOSC2 encompass 8 fields, each covering 4.1$\arcmin\times$4.1$\arcmin$. These fields were typically observed for 2.5--4 hours, with the exception of one pointing (efosc1), which was observed for 11 hours over two consecutive nights. We used the Bessel $B$ filter, since it yielded the best compromise between maximising the flux sensitivity of the CCD and minimising crowding, and an exposure time of 20 s, resulting in a cycle time of $\sim$40 s with the fast readout and additional overheads. During this first run, we were extremely lucky with the weather, not losing any time and working mostly under good seeing conditions (see Table \ref{stats} for the seeing under which the data for the different fields were obtained). Accordingly, the detection rate of EHB stars selected from the WFI/ACS photometry in the field of view monitored is quite high considering the faintness of these stars and the crowdedness of the field. For the highest quality dataset (efosc1) we detected nearly 80\% of the EHB stars, while the completeness was around 60\% for the combined sample. Please see the $N_{EHB}$ (number of EHB stars present in the field according to the ACS/WFI catalogue) and $N_{DET}$ (number of EHB stars detected on the EFOSC2 chip from that sample) columns in Table \ref{stats} for details. Note that the completeness of the sample for a given pointing depends largely on the crowding of the EHB stars in the field, as well as the seeing conditions during the run. 

For each of the EHB stars detected in the fields observed we computed light curves for all datasets available (just one for most fields, but efosc1 and efosc3 were each observed twice). Using an automated script, we first converted the relative magnitudes to relative flux, excluded any bad datapoints and outliers, and then normalised each light curve. After combining the light curves for a given star from different datasets (if applicable) we then computed the Fourier spectrum and estimated the average noise level in the 1--15 mHz range of interest (corresponding to periods of 67--1000 s; the noise level is simply the average of the Fourier spectrum in this region after taking out any pulsation peaks). At frequencies below 1 mHz the noise level increases markedly due to atmospheric effects, therefore we excluded this region. Of course, this means that our data are not sensitive to longer periodicities such as those found in the slowly pulsating sdB stars in the field. Note for completeness that we checked all targets for pulsations out to the Nyquist frequency ($\sim$25 mHz for the EFOSC2 data) but found no pulsations faster than 80 s. 

\begin{table*}[t]
\centering
\caption{Statistics for the different fields monitored} 
\begin{tabular}{ccccccccc}
\hline\hline
Field & N$_{EHB}$ & N$_{DET}$ & $Det_{\rm var}$ & $Det_{\rm lowamp}$ & N$_V$ & t(h) & seeing ($\arcsec$) & Noise (\%)\\
\hline
efosc1$^*$ & 48 & 37 & 24 (24) & 10 (13) & 2 (V1,V2) & 11.1 & 1.17 (0.9--1.9) & 0.19 \\
efosc2 & 45 & 24 & 8 (9) & 0 (0) & 0 & 3.6 & 1.21 (0.9--2.3) & 0.48 \\
efosc3 & 27 & 11 & 5 (6) & 2 (3) & 0 & 5.1 & 1.32 (1.0--2.0) & 0.35 \\
efosc4 & 52 & 21 & 16 (18) & 4 (10) & 1 (V3) & 4.2 & 1.04 (0.8--1.6) & 0.19 \\
efosc5 & 16 & 13 & 11 (11) & 1 (4) & 1 (V4) & 3.6 & 1.56 (1.2--2.6) & 0.20 \\
efosc6 & 35 & 15 & 2 (4) & 0 (0) & 0 & 3.6 & 1.50 (1.2--2.3) & 0.49 \\
efosc7 & 27 & 21 & 0 (0) & 0 (0) & 0 & 3.2 & 1.87 (1.5--2.5) & 1.34 \\
efosc8 & 63 & 40 & 6 (9) & 0 (0) & 0 & 3.3 & 1.42 (1.0--2.4) & 0.42 \\
efosc9$^{**}$ & 63 & 35 & 0 (3) & 0 & 1 (V3) & 7.8 & 1.35 (0.9--1.9) & 0.61 \\
ucam & 123 & 123 & 94 (96) & 36 (40) & 3 (V1,V2,V5) & 48.8 & 1.11 (0.7--2.6) & 0.18 \\ 
\hline 
TOTAL & & 293 & 142 (153) & 43 (57) & 5 & 94.3 \\
\hline
\hline
\end{tabular} \\
\label{stats}
$^*$ entirely overlapped by ucam field \\
$^{**}$ partly overlapped by efosc4; 10 of the N$_{DET}$ stars coincide, including V3 
\end{table*}

The number of EHB targets for which reasonable detection limits could be placed for the presence of pulsations (defined by a 3.7-$\sigma$ detection threshold of 1\%, see Section \ref{sec:stats} for details) is a function mostly of the quality of the dataset (i.e. the number of images obtained and the observing conditions). Due to the inherent difficulty of observing faint globular cluster stars from the ground using a medium-size telescope, we were able to place 1\% detection limits for pulsations on just 20\% of the EHB sample in the fields monitored (see Table \ref{stats}). Nevertheless, and much to our delight, we detected rapid multi-periodic pulsations with periods in the 84--125 s range and maximum amplitudes between $\sim$0.9 and 2.7\% in 4 H-rich sdO stars. The Fourier spectra and pulsation properties extracted from the EFOSC2 (2009) data for these four objects (hereafter V1--V4) have already been published \citep{randall2011}, and are therefore not discussed here. However, we will make use of the 2009 EFOSC2 time-series photometry sample in the statistics discussed below. 

In 2013, we returned to the NTT and gathered EFOSC2 data for one extra pointing (efosc9), selected to include a promising pulsator candidate with similar atmospheric parameters as the known pulsators (172409, see Table \ref{atmo}) as well as V3 from the 2009 field efosc4. While we used the exact same instrumental setup as for the 2009 observations, we were not so lucky with the observations the second time around. Neither the seeing nor the sky transparency were optimal, and in addition we experienced intermittent problems with the telescope focus. The completeness of the EHB sample is just 55 \%, slightly below the average for the 2009 pointings. Even though we gathered nearly eight hours of useful time-series photometry for the new field, the noise level of the data is comparable to that achieved for efosc2 or efosc6, which were observed for only three and a half hours under similar seeing conditions. To a large extent this is due to the placement of efosc9 relatively close to the cluster centre and the resulting high crowding of the field. The more crowded the field, the more the quality of the photometry is affected by bad and variable seeing. Comparing e.g. the noise level in the light curve obtained for V3 in efosc4 (seeing $\sim$1.1$\arcsec$) and efosc9 (seeing $\sim$1.4$\arcsec$) reveals the former to be a factor of 3.5 lower, even though only just over half the number of images were gathered. Nevertheless, we were able to confirm the two known pulsation periods of V3 from the new data at 109.8 s (with a relative amplitude $A$=3.8\%) and 103.1 s ($A$=2.3\%). Unfortunately, no new pulsators were discovered, and the detection limits we were able to place on the EHB stars detected were modest (including for the promising candidate variable), but the results are included in the statistics below.

\subsection{Follow-up observations with ULTRACAM}

The aim of the ULTRACAM observations was to secure a longer time-series for just one field containing the previously identified variables V1 and V2 in order to better characterise the pulsation spectrum and eventually perform asteroseismology. At the same time, a relatively large (6$\arcsec\times$6$\arcsec$) field would be monitored in depth, providing high-quality lightcurves for over 100 EHB stars and the chance to discover relatively low-amplitude pulsators. The ULTRACAM pointing completely covers the previously monitored field efosc1, as well as a region of $\omega$ Cen that had not been monitored before. We were allocated a total of 6 nights on the NTT, one of which was lost due to weather. Over the remaining 5 nights, we gathered nearly 50 hours of simulteneous multi-colour $u'g'r'$ data with an exposure time of 6 s. ULTRACAM is extremely efficient for time-series photometry as there is negligible deadtime (25 ms) in between exposures, and the cycle time is effectively the same as the exposure time.

While the ULTRACAM data were obtained in three bands (Sloan $u'g'r'$), only the $u'$ data are used for the statistics presented here. Both the $g'$ and particularly the $r'$ data suffered from severe crowding, especially during periods of bad seeing, and this combined with the lower intrinsic flux of the EHB stars at longer wavelengths renders the $r'$ data completely useless. The $g'$ data were found to be useful for individual EHB stars that are subject to less crowding, but overall their inclusion degraded the noise level and detection thresholds achieved. Therefore, we focus on the more reliable u' data. We computed the combined nightly ULTRACAM light curves, Fourier spectra and detection thresholds following the exact same procedure as for the EFOSC2 data. Note that again we searched the Fourier spectra out to the Nyquist frequency, but found no pulsations with periods less than $\sim$ 80 s despite being sensitive to periods as low as $\sim$ 12 s.

As expected, the quality of the ULTRACAM $u'$ data is significantly higher than that achieved on average for the EFOSC2 fields. This is due to the much longer time spent monitoring the targets, as well as the $u'$ optimised ULTRACAM CCD and the nearly 100\% duty cycle. For details please see Fig. \ref{noise}, where we show a histogram of the noise level in the Fourier spectrum for each of the light curves extracted from the EFOSC2 fields compared to that for the ULTRACAM field. All of the 123 ACS/WFI EHB stars in the field were detected in at least some of the ULTRACAM $u'$ images, and a reasonable (1\%) detection threshold was achieved for 76\% of these, allowing us to derive some meaningful statistics on the pulsator fraction among EHB stars in $\omega$ Cen. In addition, we discovered a new EHB variable, V5, that exhibits lower amplitude pulsations than the sdO variables found from the EFOSC2 data (see Section \ref{vars}).

\begin{table*}
\caption{Details of the $\omega$ Cen variables and the slower variable (SV). The periods for V3 and V4 are taken from \citet{randall2011}.}
\centering
\begin{tabular}{l l l l l l l l}
\hline\hline
{} & $\alpha$(2000.0) & $\delta$(2000.0) & $B$ & $T_{\rm eff}$ (K) & Periods (s) \\
\hline
V1 & 13:27:11.8 & $-$47:32:29 & 18.30 & 48,500 & 115.0, 84.7 \\
V2 & 13:27:04.8 & $-$47:32:32 & 18.44 & 49,900 & 101.7, 107.8 \\
V3 & 13:26:24.7 & $-$47:24:52 & 18.30 & 49,300 & 109.9, 102.9 \\
V4 & 13:26:15.0 & $-$47:33:06 & 18.21 & 52,000 & 123.6, 113.5 \\
V5 & 13:27:15.0 & $-$47:30:51 & 18.51 & 53,400 & 100.6, 99.2, 107.5 \\
\hline \\
SV & 13:26:59.2 & $-$47:30:27 & 18.01 & 48,200 & 2685 \\
\hline \\ 
\end{tabular}
\label{variables}
\end{table*}

\begin{figure*}[t]
\centering
\begin{tabular}{ccc}
{\includegraphics[width=6.0cm,angle=0,bb=0 0 580 500]{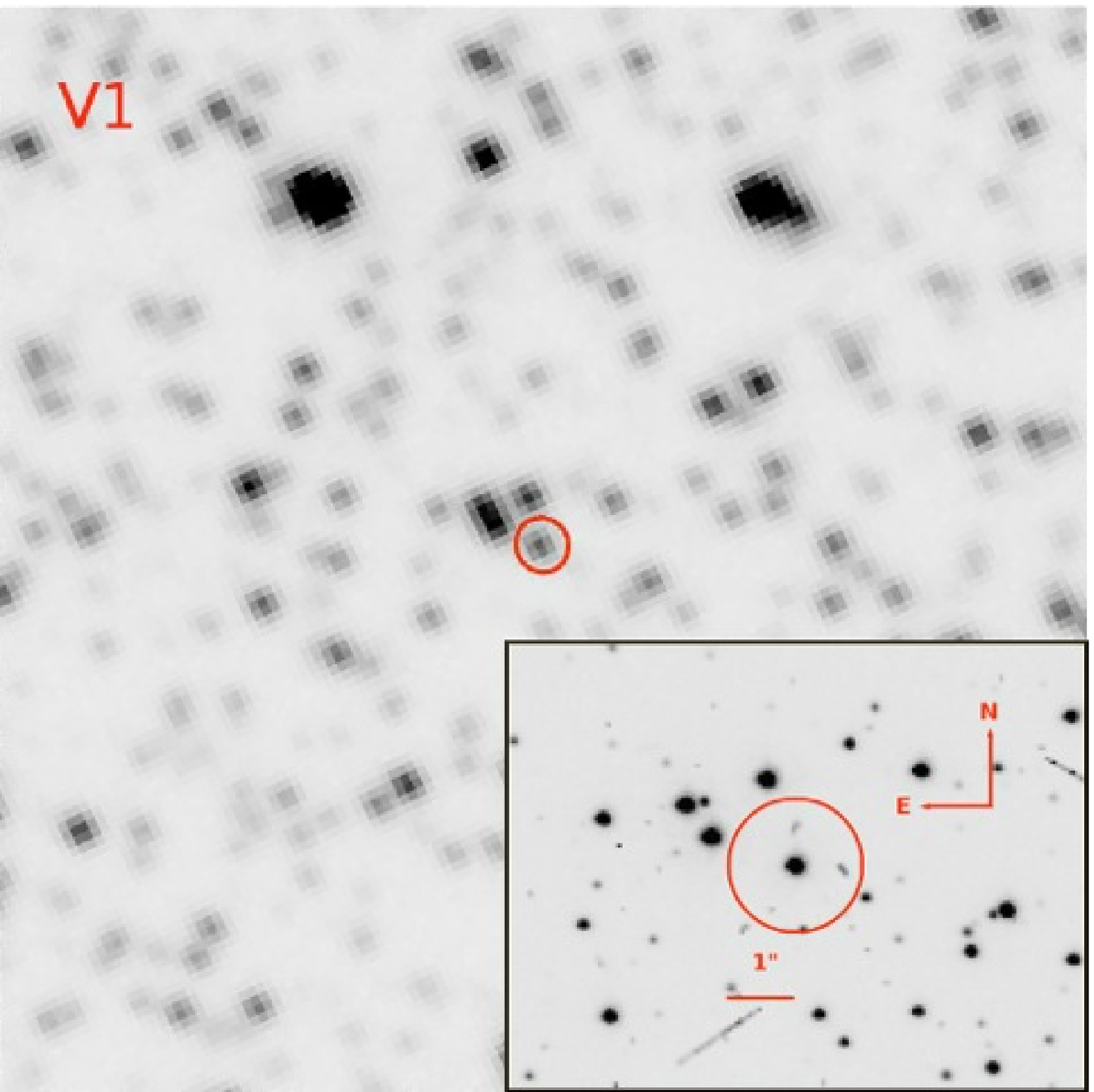}} & {\includegraphics[width=6.0cm,angle=0,bb=0 0 580 500]{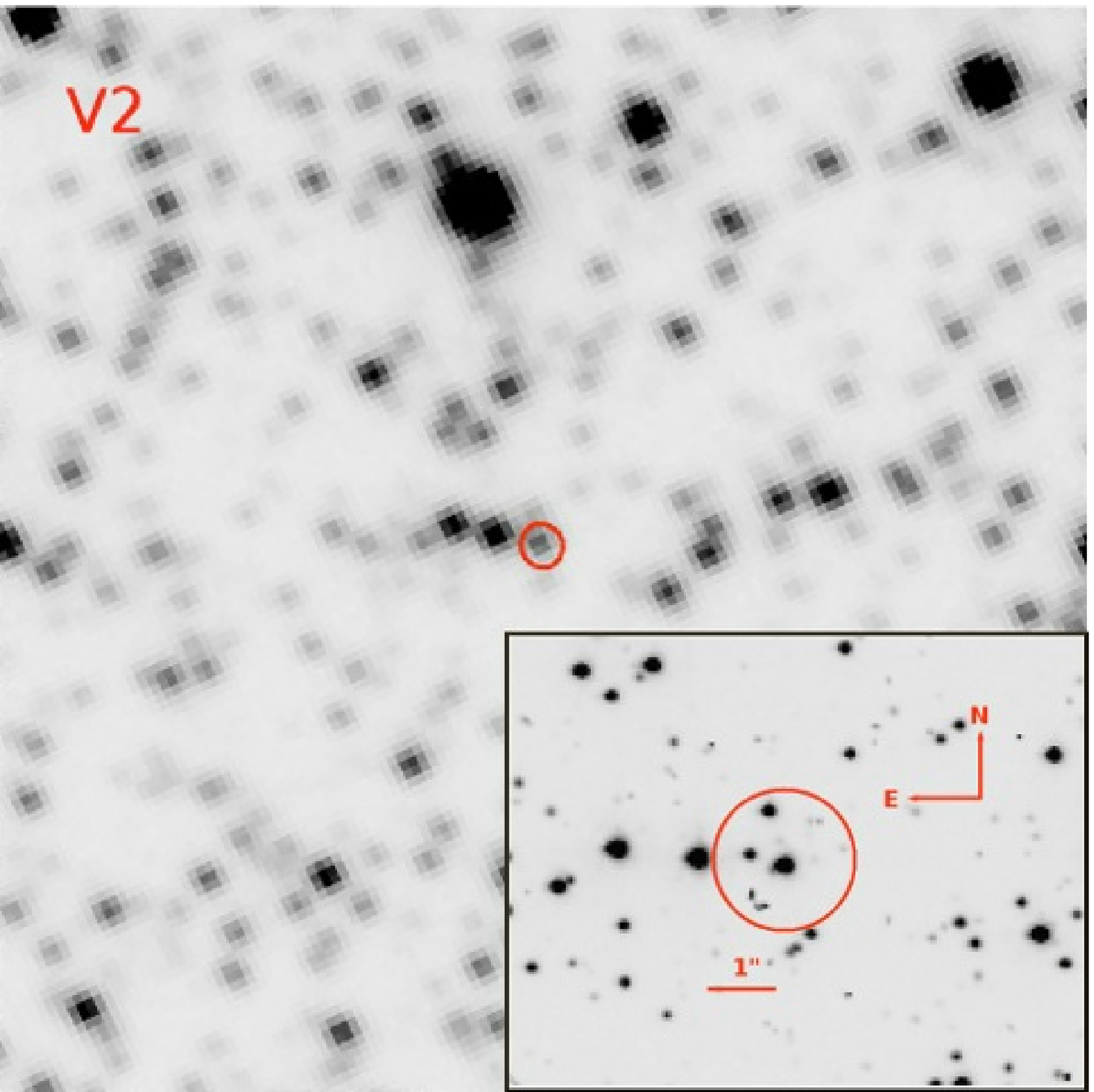}} & {\includegraphics[width=6.0cm,angle=0,bb=0 0 580 500]{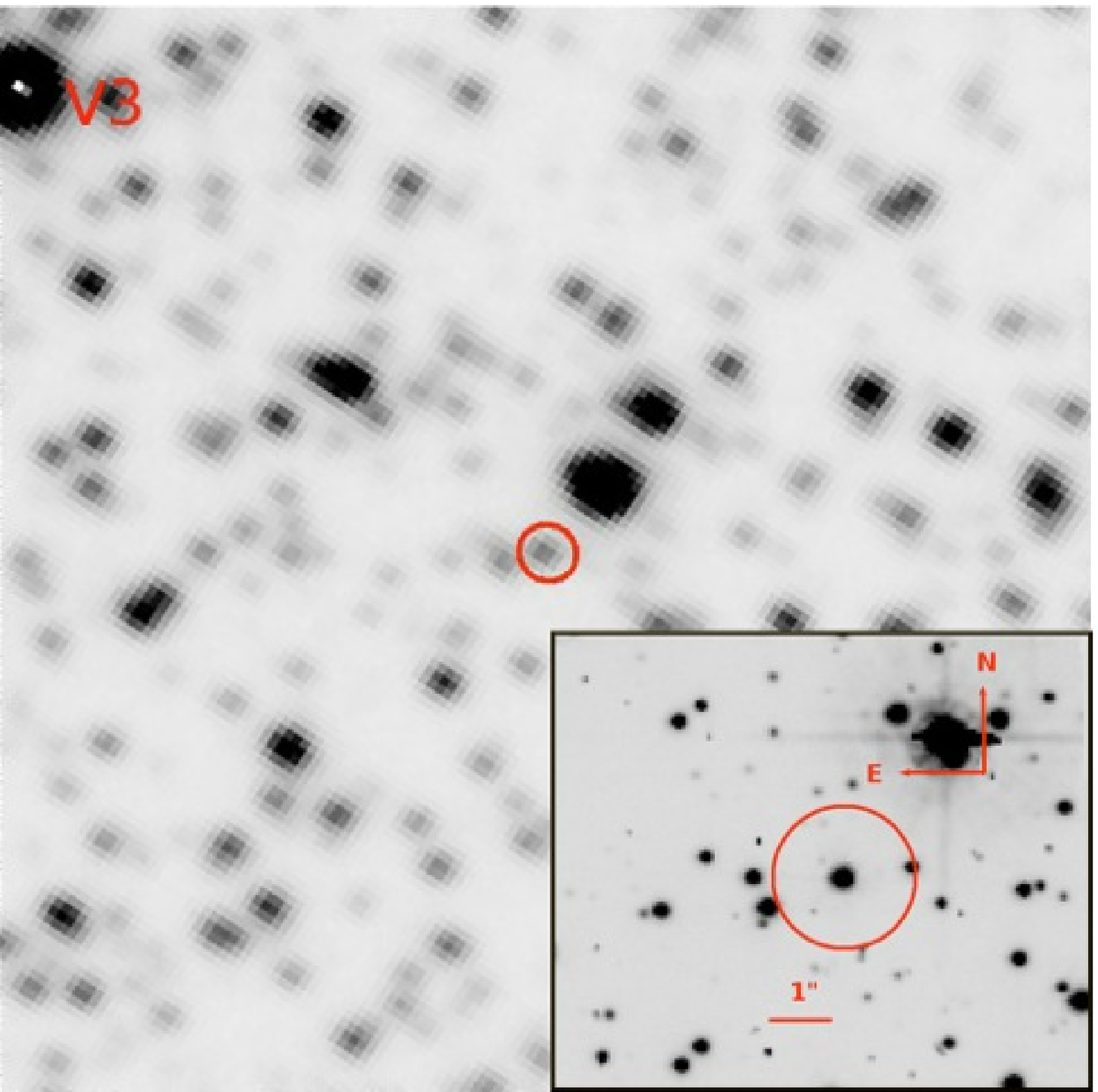}} \\
{\includegraphics[width=6.0cm,angle=0,bb=0 0 580 500]{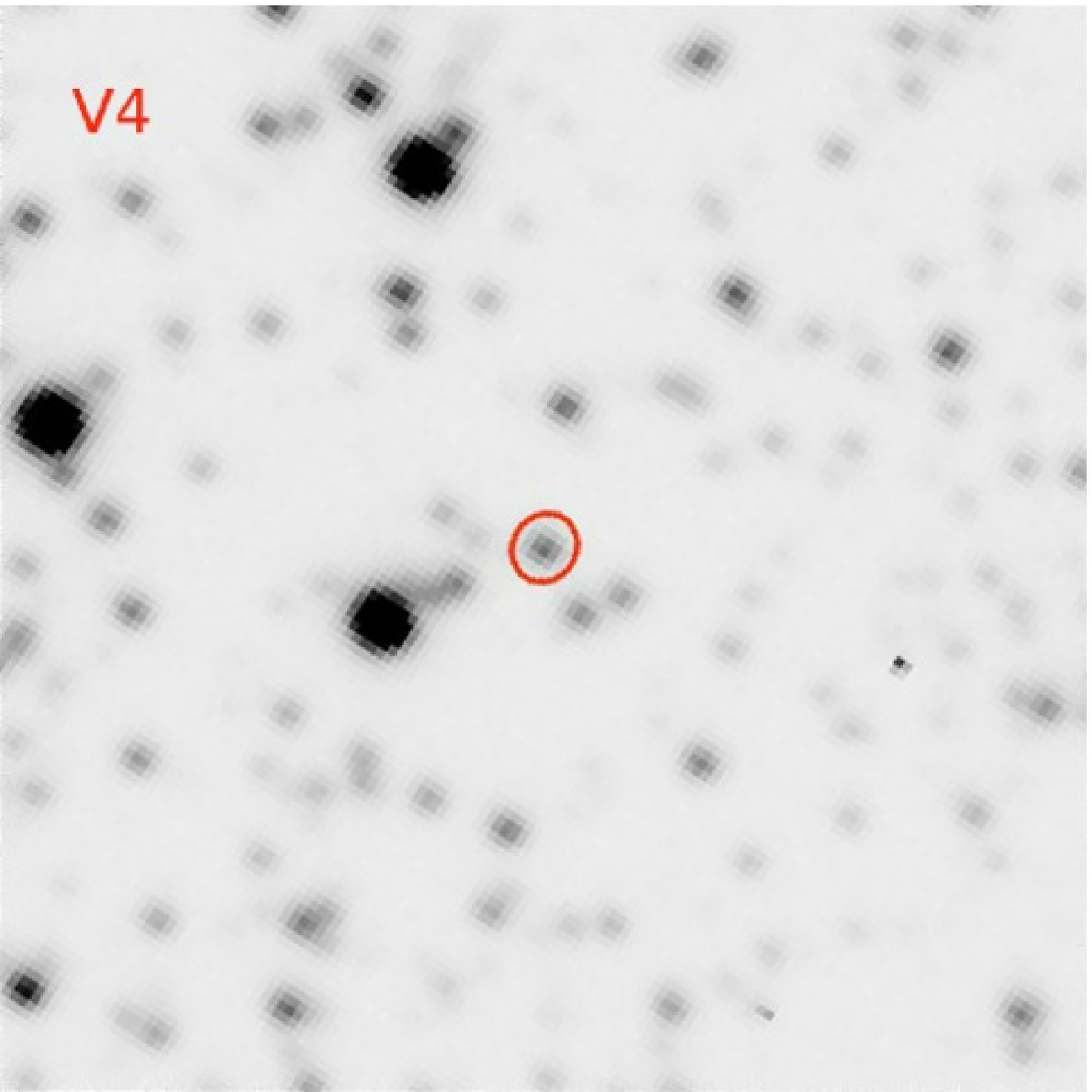}} & {\includegraphics[width=6.0cm,angle=0,bb=0 0 580 500]{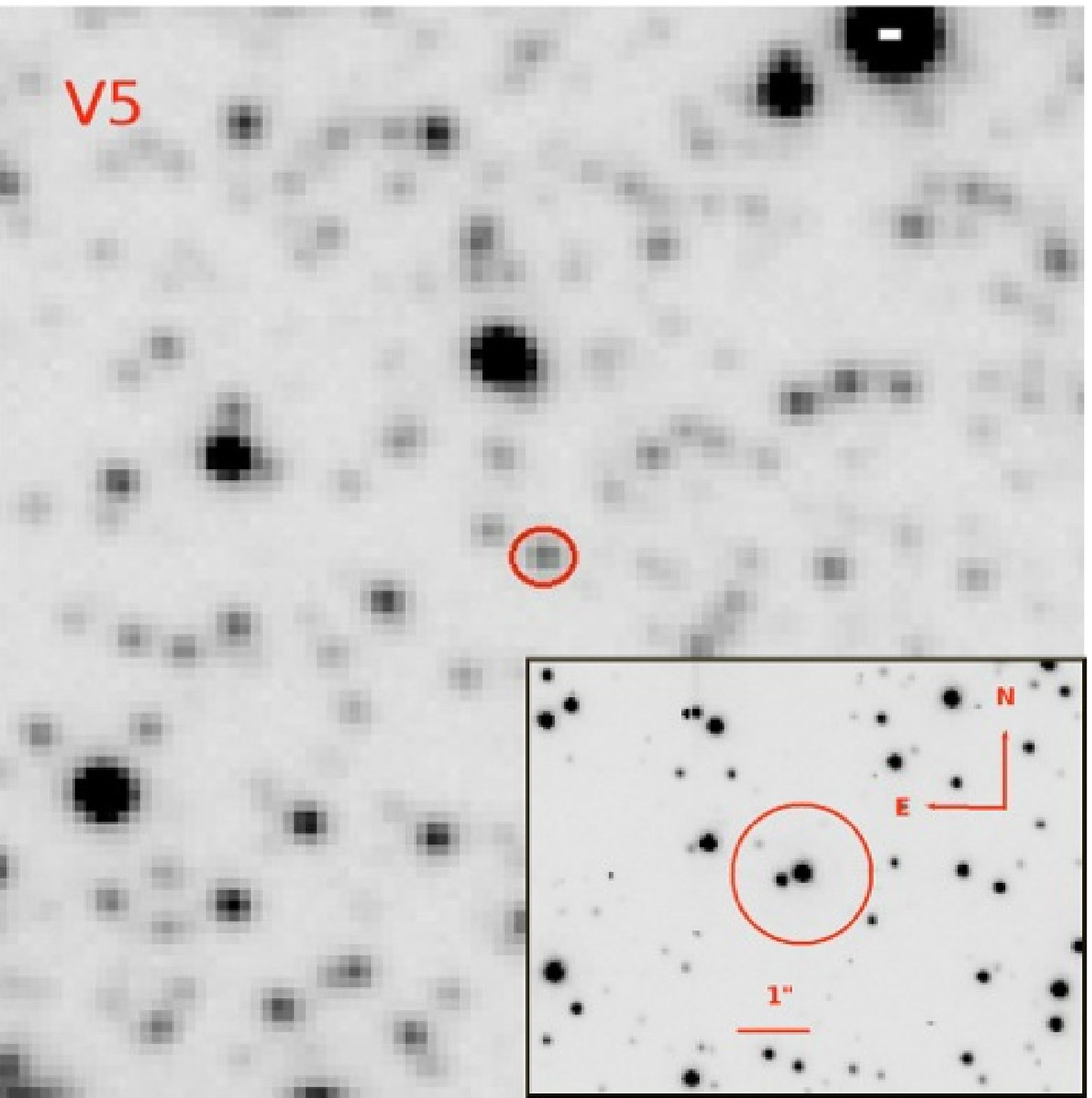}} & {\includegraphics[width=6.0cm,angle=0,bb=0 0 580 500]{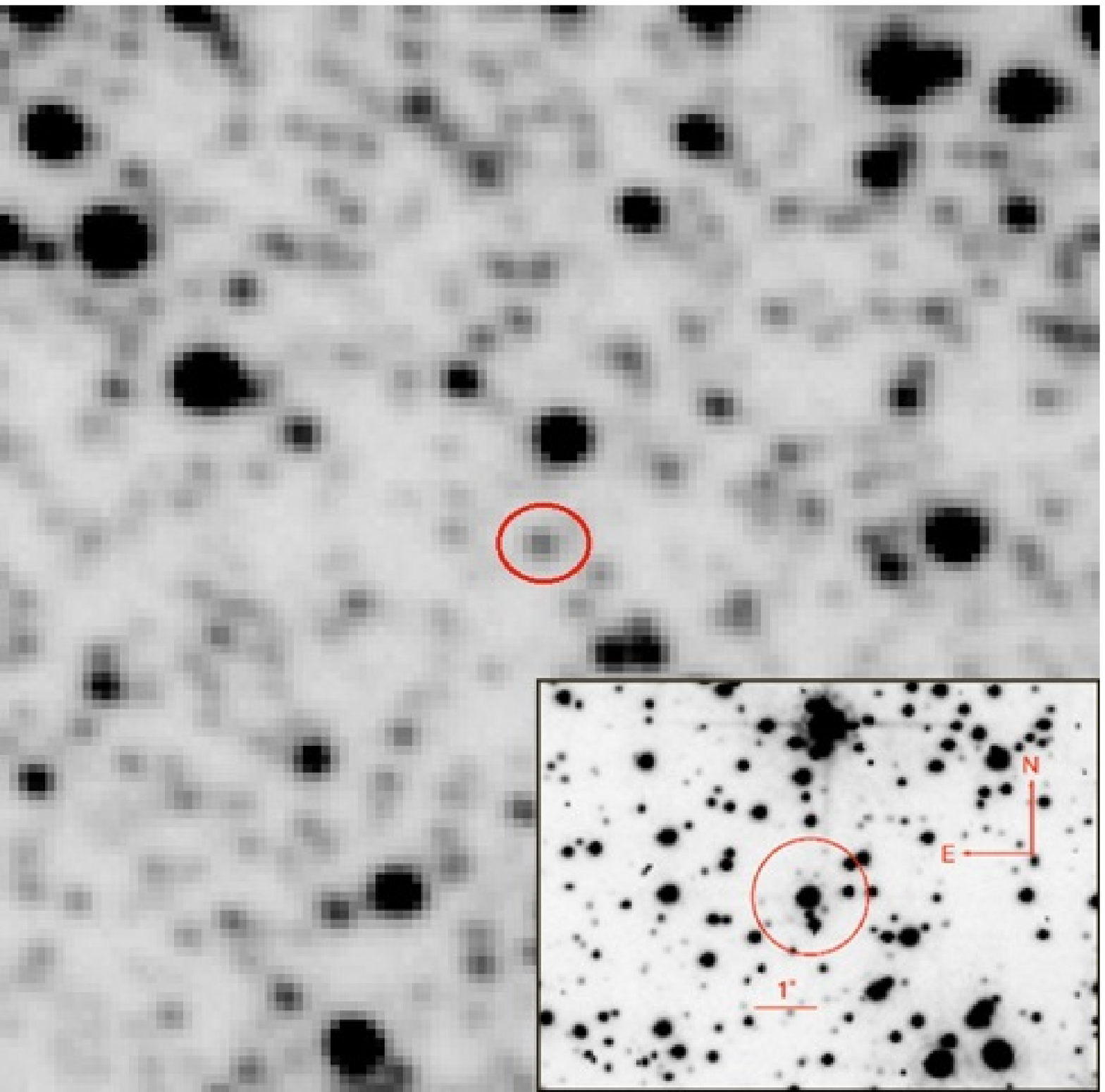}}
\end{tabular}
\caption{Finding charts for the five $\omega$ Cen variables labelled V1--5 as well as the slower variable (bottom right) based on FORS2 preimaging taken under $\sim$0.8$\arcsec$ seeing in the Johnson $B$ band. The size of each image is 30$\arcsec\times$30$\arcsec$ with North at the top and East to the left. The coordinates of the respective pointings can be found in Table \ref{variables}. The inlay in the bottom right-hand corner shows a zoomed-in finding chart based on ACS F435W images, where available.}
\label{findingcharts}
\end{figure*}

\subsection{Statistics from the EFOSC2/ULTRACAM time-series photometry}
\label{sec:stats}
 
In order to derive the most meaningful statistics possible we pooled the light curves from the EFOSC2 and the ULTRACAM fields. A summary of the parameters of interest is given in Table \ref{stats}, where for each field we list the number of EHB stars present in the field according to the ACS/WFI catalogue (N$_{EHB}$), the number of EHB stars that were detected in at least 100 EFOSC2/ULTRACAM images (N$_{DET}$), the number of targets for which the 3.7$\sigma$ detection threshold is less than 1.0 \% ($Det_{\rm var}$), the number of targets for which the 3.7$\sigma$ detection threshold is less than 0.5 \% ($Det_{\rm lowamp}$), the number of variables detected ($N_V$), the time spent observing the field $t$, the median seeing together with the range of seeing encountered, and the median noise level in Fourier space of all light curves extracted from the data. The two detection parameters $Det_{\rm var}$ and $Det_{\rm lowamp}$ were chosen as reasonable thresholds above which one may hope to detect a) high-amplitude pulsations similar to those measured for the $\omega$ Cen sdOs V1-V4 and some of the higher amplitude rapid sdB pulsators ($Det_{\rm var}$), and b) low-amplitude pulsations typical of more normal sdB pulsators ($Det_{\rm lowamp}$). The corresponding values in brackets give the number of stars that do not show peaks in the Fourier spectrum above these (1.0 and 0.5 \%) limits. This latter category contains all the targets with 3.7$\sigma$ noise levels below the limits, as well as objects with higher noise levels but where the highest (spurious) peaks are below the limits. In some ways it is the more interesting value, as it gives the number of stars for which we can {\it exclude pulsations} down to the quoted level.   

\begin{figure*}[t]
\centering
\includegraphics[width=15cm,angle=0]{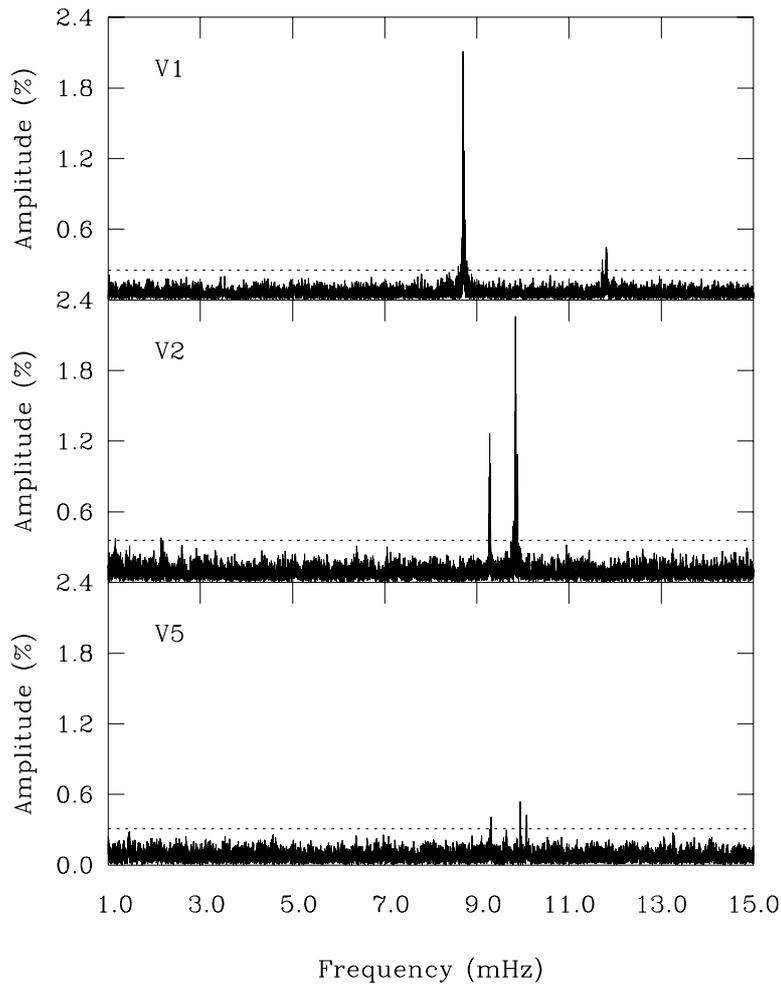}
\caption{Fourier amplitude spectra for V1,V2 and V5 based on the combined ULTRACAM $u'$ light curve. The horizontal dashed line indicates the 3.7-$\sigma$ detection threshold for each data set.
}
\label{fts}
\end{figure*}  

\begin{table*}[t]
\caption{Periodicities extracted for V1,V2 and V5 on the basis of the ULTRACAM $u'$ data.}
\centering
\begin{tabular}{l l l l l}
\hline\hline
{} & Period (s) & Freq. (mHz) & A (\%) & Freq. components \\
\hline\hline
V1 \\
\hline
1 & 115.0 & 8.70 & 2.4 & 114.98 and 114.92 s peaks \\
2 & 114.7 & 8.72 & 2.2 & 114.69, 114.64 and 114.57 s peaks \\
3 & 84.7 & 11.81 & 0.60 & 84.73, 84.59 s peaks \\ 
4 & 114.4 & 8.74 & 0.47 & \\
5 & 84.3 & 11.86 & 0.36 & \\
\hline
V2 \\
\hline
1 & 101.7 & 9.83 & 2.5 & 101.68 and 101.63 s peaks \\
2 & 107.8 & 9.28 & 1.2 & 107.78 and 107.77 s peaks \\
3 & 101.2 & 9.88 & 1.4 & 101.27 and 101.22 s peaks \\
\hline
V5 \\
\hline
1 & 100.6 & 9.94 & 0.54 & \\
2 & 99.3 & 10.08 & 0.45 & \\
3 & 107.5 & 9.30 & 0.42 & \\
\hline
\end{tabular}
\label{freqs}
\end{table*}

In the computation of the total number of EHB stars detected from our observations as well as the associated detection thresholds we take into account the overlaps between different pointings. The field efosc1 for example is ignored here since it is completely overlapped by the higher quality ucam field. Similarly, efosc4 partially overlaps with efosc9. For the 10 detected EHB stars that coincide for the two fields the efosc4 light curves are all of higher quality, so it is these that are used in the statistics. Note that we attempted to combine the light curves obtained for the same stars from different pointings, but this degraded the quality of the Fourier spectrum.

Strictly speaking, the ULTRACAM and EFOSC2 data show systematic amplitude differences at the quantitative level since they were obtained with slightly different passbands. All other things being equal, the measured amplitudes of the pulsations as well as the noise level (which is dominated by differential atmospheric variations) are expected to be comparatively larger in the $u'$ compared to the $B$ by as much as $\sim$ 20--30\%. Therefore, the effective noise level achieved for the ULTRACAM pointing is in fact quite a bit better than that for efosc1, despite the values in Table \ref{stats} being very similar. At the qualitative level and in terms of the number of stars for which detection limits can be placed however the data are comparable. Note also that the pulsation periods are independent of the bandpass used for observation.   

From Table \ref{stats} it can be seen that from the data gathered we could have potentially detected pulsations at the 1\% level for some 150 EHB star candidates. Given that we actually found 4 pulsators at that amplitude level, the fraction of high-amplitude variables compared to EHB candidates in our sample is $\sim$2.7\%. At the lower detection threshold of 0.5\%, we would have detected pulsations for 57 EHB star candidates, which taking into account the 5 variables detected (the four high-amplitude ones plus one at lower amplitude) gives a pulsator fraction of $\sim$10 \%. Note that all the five variables detected are H-rich sdO stars, rather than analogs of the sdB stars known to pulsate among the Galactic field population. While our observations would not have been sensitive to the long-period sdB variables, we should have detected any typical sdBV$_{r}$ stars among the 57 targets where we achieved a detection threshold less than 0.5\%. If we look purely at this last result and assume an sdBV$_r$ pulsator fraction of 5\% for the field, we would have expected to find about 2--3 sdBV$_r$ stars in $\omega$ Cen in the course of our survey. Does this mean that rapidly pulsating sdB stars do not exist in $\omega$ Cen? In order to address this question, we need to take into account the distribution of our selected EHB candidates in $\log{g}$-$T_{\rm eff}$ space, which we do in Section \ref{spec}.

\subsection{The $\omega$ Cen variables}
\label{vars}

The combined EFOSC2/ULTRACAM time series photometry revealed a total of 5 rapid non-radial pulsators in $\omega$ Cen. Four of these, V1--V4, were found on the basis of the EFOSC2 data alone, and the corresponding Fourier spectra have already been published \citep{randall2011}. Here, we show the new ULTRACAM data for V1 and V2, and also present a new variable, V5. The coordinates and main characteristics of all five $\omega$ Cen pulsators are listed in Table \ref{variables} and they are highlighted in the CMD shown in Fig \ref{cmd}. We also include finding charts in Fig. \ref{findingcharts}. Given the crowdedness of the field these will be useful to anybody planning follow-up observations on these targets.

The Fourier amplitude spectra for the three variables observed with ULTRACAM $u'$ are depicted in Fig. \ref{fts}. Based on these data, we attempted to extract all periodicities present down to a threshold of 3.7 times the mean noise level using standard pre-whitening techniques. This was straightforward to do only for V5, which exhibits rather low-amplitude pulsations that show no obvious fine structure down to the detection threshold. For the higher amplitude variables V1 and V2 on the other hand, the pre-whitening procedure revealed extremely complex fine structure around the main peaks visible in Fig. \ref{fts}. For V1, the splitting of the dominant peak was already evident from the EFOSC2 data, however the split components were spaced further apart than in the ULTRACAM data \citep[cf Table 1 from][]{randall2011}. This together with the fact that some of the split components are separated by a frequency similar to the resolution of the data set (2.1 $\mu$Hz) is taken to imply that the splitting is at least partially an artefact, likely caused by amplitude variations in the pulsations. And indeed, the pulsations in both V1 and V2 appear to have undergone significant amplitude variations when comparing the EFOSC2 and the ULTRACAM data. For instance, the previously dominant 115.4 s periodicity in V2 has now completely vanished, as has the 119.1 s peak in V1. Note that such amplitude variations are very common among the field sdBV$_r$ stars \citep[see, e.g.][]{kilkenny2010}, and are therefore not unexpected.

\begin{figure*}[t]
\centering
\begin{tabular}{cc}
{\includegraphics[width=7.5cm,angle=0,bb=150 200 550 650]{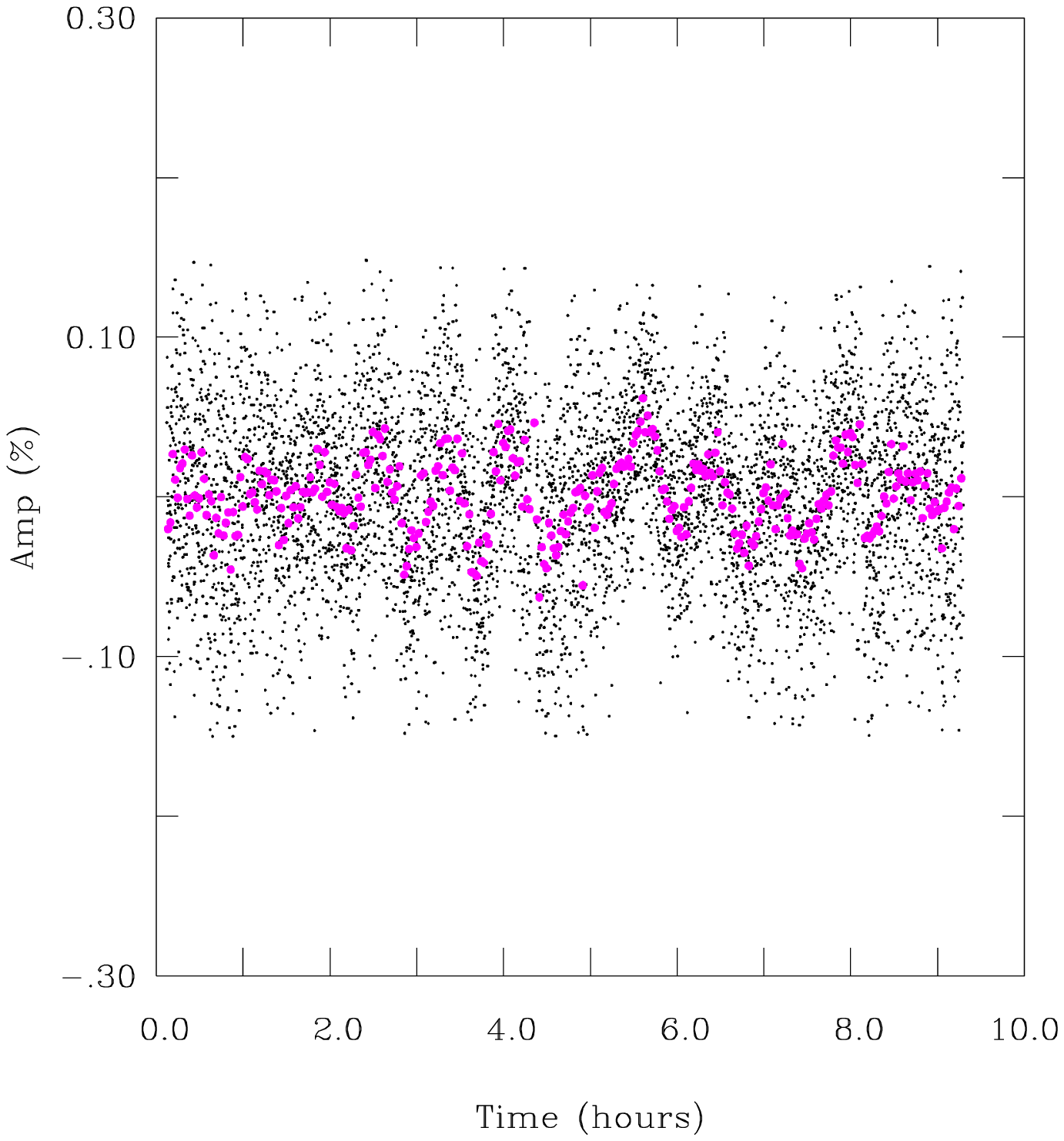}} & {\includegraphics[width=8.0cm,angle=0,bb=80 205 500 650]{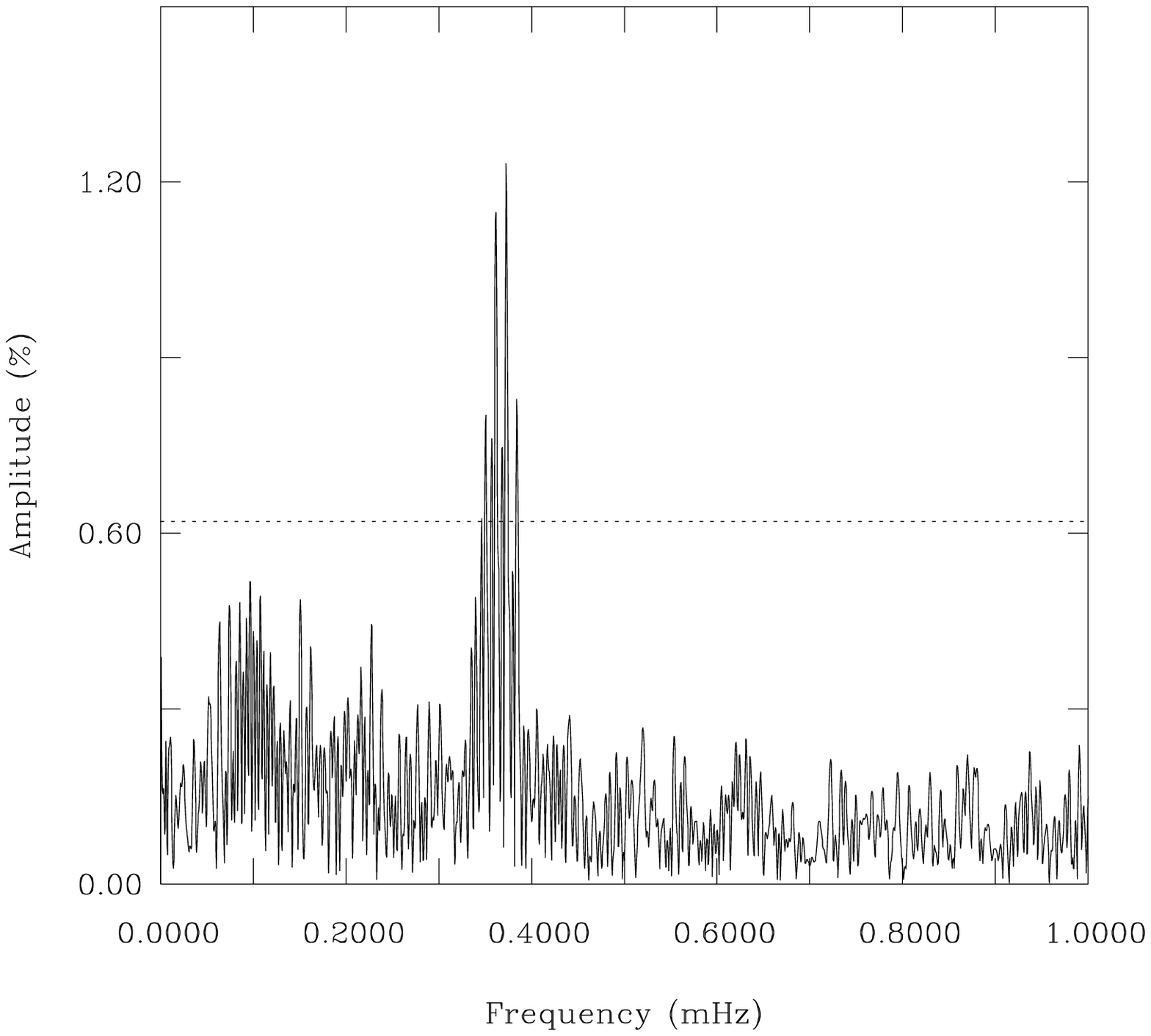}}
\end{tabular}
\caption{{\it Left:} Light curve for the EHB star 248955 from April 25th. A low-order polynomial was fit to remove differential extinction over the night, and outliers were removed. The black points represent the original ULTRACAM $u'$ data with a sampling time of 6 s while the magenta points have been binned to 100 s to reduce the noise and reveal the long-period luminosity variations more clearly. {\it Right:} Fourier spectrum of the combined light curve from 22--25 April in the 0--1 mHz range. The horizontal dashed line indicates 4 times the mean noise level, above which the excess power around 0.34 mHz is clearly visible.}
\label{ell}
\end{figure*}   

For the periodicities reported in Table \ref{freqs} we grouped together the finely split frequency components extracted from the light curve where applicable, and added the amplitudes of the individual components to give the total amplitude of the combined peak. Note that there is still some splitting of the main peaks, albeit at a frequency spacing one order of magnitude larger than the resolution of the dataset. While we list these frequencies separately in Table \ref{freqs} for completeness, we do not believe they constitute separate harmonic oscillations. Therefore, only the highest amplitude component is listed in the summary table for the variables (Table \ref{variables}).

Overall, the outcome of the period analysis of the ULTRACAM data for V1 and V2 is slightly disappointing. While the aim had been to uncover several extra frequencies beyond those measured from the EFOSC2 survey and eventually attempt asteroseismology, the stars simply did not cooperate. It appears that in general the $\omega$ Cen variables are dominated by just a few (2--3) pulsations, any other periodicities being too weak to be detected even from the ULTRACAM data. It is difficult to see how one can obtain much better period spectra using current observational facilities and taking into account the amount of time typically allocated on large ground-based telescopes or space facilities. For the time being, it appears that quantitative asteroseismology is out of reach for these stars, since this typically requires at least 5--10 observed frequencies. Therefore, we concentrate our efforts on the qualitative characterisation of the pulsation properties of the $\omega$ Cen EHB variables in what follows.

\subsection{Slower luminosity variations}

The time-series photometry obtained both with EFOSC2 and ULTRACAM was geared towards detecting short-period luminosity variations on the order of tens to hundreds of seconds. And indeed, a visual inspection of the EHB light curves reveals that on longer time scales of thousands of seconds and more the light curves are generally dominated by atmospheric variations, and the detection thresholds in the Fourier domain become prohibitively large. We nevertheless decided to run a quick check for any obvious periodicities at lower frequencies. 

Somewhat unexpectedly, we noted an apparent periodicity around 2700 s in the light curve for one of our targets (see Fig. \ref{ell}). The Fourier spectrum reveals a strong peak at 2684.6 s (0.34 mHz) with an amplitude of 1.2\%. There are no further credible peaks above the detection threshold, which was set to four times the mean noise level (compared to the 3.7 $\sigma$ threshold adopted for the higher frequency regime discussed in the previous sections this more conservative threshold somewhat compensates for the noise increase towards the low-frequency end of the frequency range considered here). Note that none of the other EHB stars monitored show similar variations, indicating that they are intrinsic to this target rather than due to atmospheric effects. By sheer luck this interesting star is also part of our spectroscopic sample (see Table \ref{atmo}), thanks to which we can classify it as a H-rich sdO star with a temperature around 48,000 K. Incidentally, these properties are very similar to those of the rapid variables V1--V5 discussed above.

We find three possible explanations for the observed luminosity variation:
\begin{itemize}
\item  it is the signature of an ellipsoidal variation or possibly a reflection effect associated with close binarity. In the first scenario, the EHB star undergoes a deformation as it orbits a close, massive (likely a white dwarf, but possibly a neutron star or black hole) companion, giving rise to a very regular luminosity variation at half the orbital period. Such ellipsoidal variations have been inferred for several field EHB stars, e.g. a 4100 s ellipsoidal variation with an amplitude of 1.4 \% was observed for the sdBV$_r$ star KPD 1930+2752 \citep{billeres2000}. While the $\sim$2700 s luminosity variation observed for the target in $\omega$ Cen is quite a bit shorter than this, the implied orbital period of 5369 s (or 0.062 d) is tantalisingly close to the shortest-period sdB + WD binary known in the field \citep[CD$-$3011223 with a period of 4232 s, see][]{geier2013}. If on the other hand the observed luminosity variation were caused by a reflection effect, the implied orbital period would be equal to the photometric periodicity, i.e. 0.031 d. Reflection effects are observed when the hot subdwarf primary in a close binary periodically heats up the observable side of its low-mass, cool companion \citep[usually a dwarf M star or a brown dwarf, see e.g.][]{kupfer2015}. The shortest-period reflection effect system known \citep[PG 1621+476,][]{schaffenroth2014} has a period of 0.070 d, more than twice as long as the luminosity variation detected here, and a comparatively higher photometric amplitude of $\sim$5\%. Therefore, we believe the ellipsoidal variable scenario to be more plausible for the time being.
\item it constitutes a $g$-mode pulsation. Pulsations on time-scales of thousands of seconds are indeed quite commonly observed among sdB stars in the field. These occur for much cooler stars with temperatures below $\sim$30,000 K, and also at lower amplitudes \citep[well below 0.5\%, see e.g.][]{green2003}. However, given that the rapid pulsators in $\omega$ Cen are found at much higher temperatures than similar variables in the field it is conceivable that longer periods might also be excited at higher temperatures. In this case we would expect to see several periodicities. Unfortunately, the quality of our lightcurve is not high enough to tell whether the luminosity variations observed are mono-periodic or not. 
\item it is not inherent to the EHB star, but caused by a contaminating (presumably cooler) star in the very crowded field. While from the FORS2 finding chart show in Fig. \ref{findingcharts} the star appears to be isolated, the ACS image reveals the presence of a close neighbour (and indeed, the FORS2 spectrum is polluted by a cooler star).  
\end{itemize}

\begin{figure*}[t]
\centering
\begin{tabular}{cc}
{\includegraphics[width=7.5cm,angle=270]{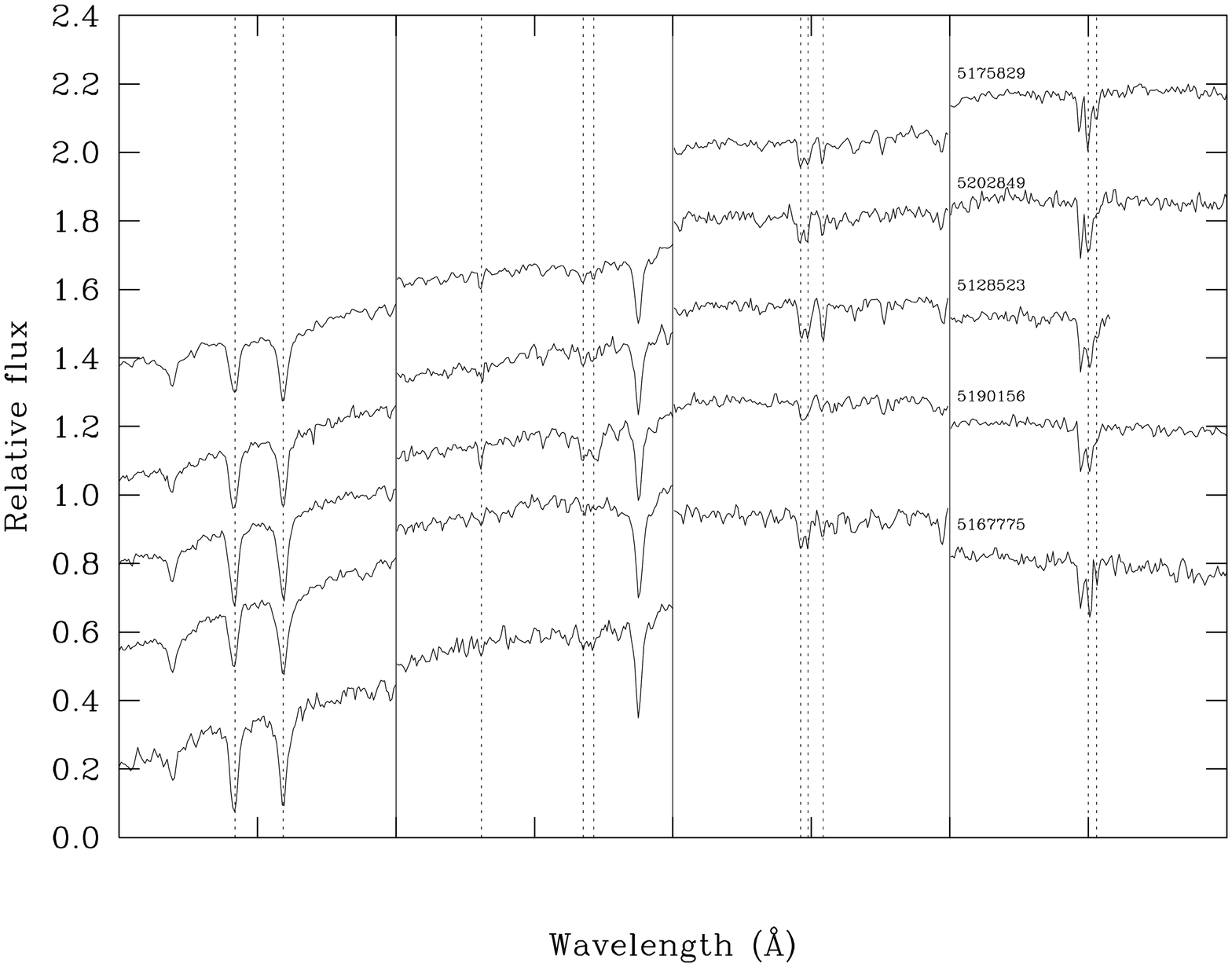}} & 
{\includegraphics[width=7.5cm,angle=270]{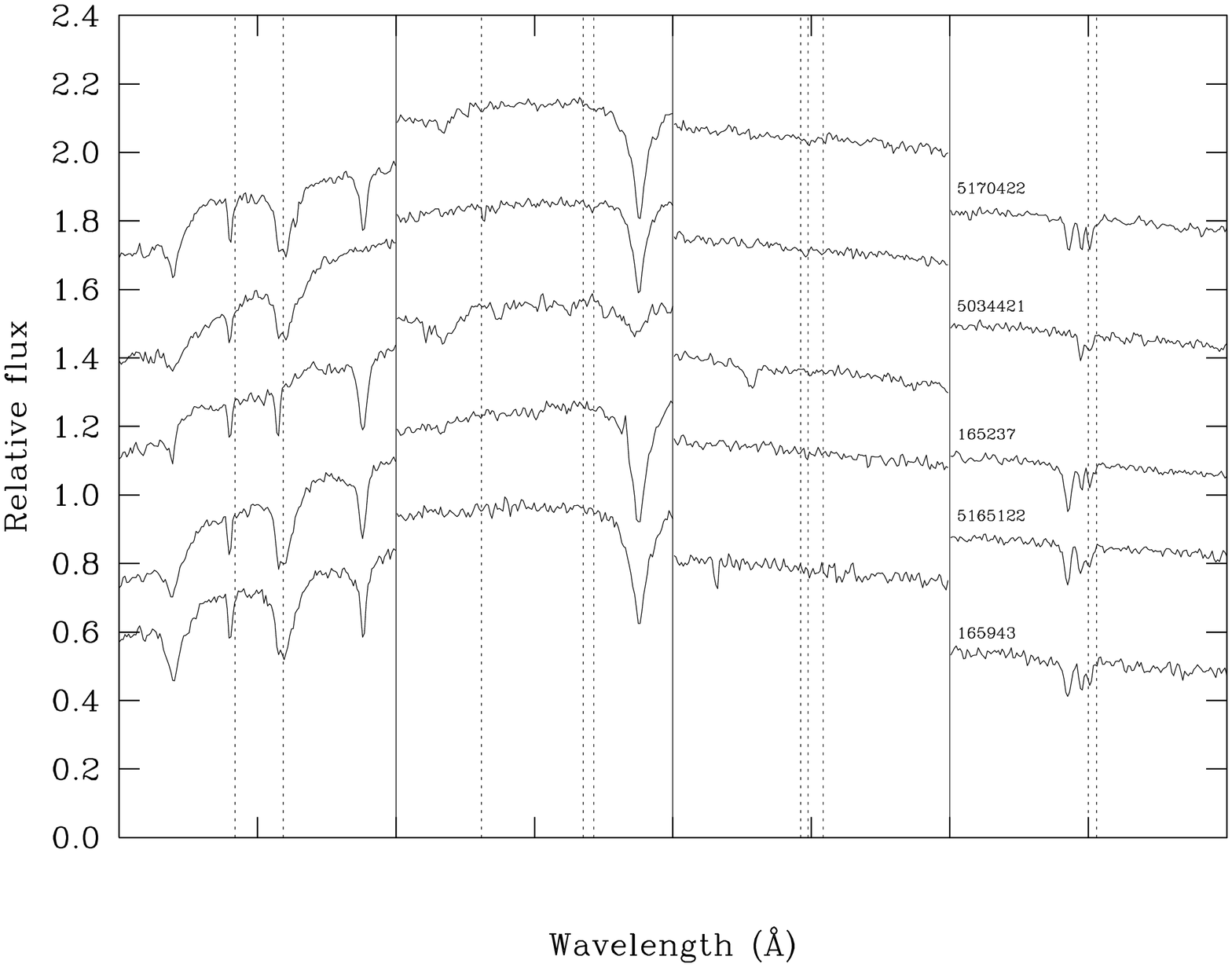}}
\end{tabular}
\caption{Some of the strongly polluted (left) and some of the ``clean'' (right) spectra present in our sample. The four panels in each figure are centred on the Ca II doublet (3830--4030 \AA\ wavelength range shown, left panel), the G-band (4165--4365 \AA, second panel from left), the Mg I triplet (5075--5275 \AA, second panel from right) and the Na I doublet (5790--5990 \AA, right panel). The dashed vertical lines indicate the wavelengths in the rest frame of the stellar spectrum of (from left to right) the Ca II K and H line, the Ca I, CH and Fraunhofer G lines making up the G-band, the Mg I triplet and the Na I doublet (see text for details). In the polluted spectra all of these lines are seen in the rest frame of the stellar spectrum, while in the ``clean'' spectra the G-band and Mg I triplet are absent and the Ca II and Na I doublets are blue-shifted, indicating an interstellar origin.    
}
\label{gpol}
\end{figure*}

Given the limited data currently available, all explanations remain viable options. The most exciting is the idea of a close binary system, since up until now no EHB binaries have been confirmed in any globular cluster with the possible exception of an unusual MS/EHB binary in NGC 6752 \citep{monibidin2015}. We plan to obtain follow-up time-series radial velocity observations in order to check this possibility. While the two available FORS spectra for this star show no discernable relative velocity shifts or obvious smearing, this is not too surprising considering the low ($\sim$150 km/s) velocity resolution and the long exposure time, and we will need to wait for better data to pursue the matter further.     

\section{The spectroscopic sample}
\label{spec}

The main aim of the spectroscopic observations obtained was to determine the atmospheric parameters of the variables identified from the photometry. At the same time, we observed as many other EHB stars as possible within the chosen fields of view, making full use of the multi-object capability of the instrument. This resulted in a very nice spectroscopic sample of EHB stars in $\omega$ Cen that is interesting in its own right. A subset of this sample was exploited by \citet{latour2014}, who performed a detailed abundance analysis focusing on carbon and helium. We refer the interested reader to that publication for more details on the atmospheric composition of the EHB stars observed, and the implications this has in the context of understanding their evolutionary history. Here, we concentrate on the pulsators and the characterisation of the instability strip in terms of effective temperature, surface gravity and helium abundance.

The spectroscopic data were obtained in service mode in March 2011 and April 2013 using the MXU mode (600B grating, 0.7$\arcsec$ slit width)  of FORS2 at the VLT on Cerro Paranal, Chile. In total, we observed three pointings that largely overlap with the fields monitored with time-series photometry, obtaining two 2750 s exposures for each. The spectroscopic targets were selected from the EHB sample described in section 2 to fit within the spatial constraints of the MXU mask and be relatively isolated on the CCD. Nevertheless, due to the crowdedness of the field several of the resulting spectra turned out to be too contaminated by neighbouring stars for a fruitful analysis (see below). Apart from the pulsators, which were of course specifically included, the EHB targets in the spectroscopic sample should not suffer from an observational bias compared to the photometric sample and can therefore be used for statistical comparisons.   

Data reduction was performed using a combination of the FORS2 pipeline and a customised IRAF procedure. The pipeline performed bias subtraction, flatfielding and wavelength calibration, then the resulting products were used as a basis to manually identify and interactively extract the relatively faint EHB target spectra with IRAF applying background subtraction. The resulting spectra were cleaned from cosmic rays and combined (2 spectra for each target)\footnote{The reduced spectra are available from S. Randall upon request.}. They have a wavelength resolution of $\sim$2.6 \AA\ and nominally cover the 3400--6100 \AA\ range, however some are truncated due to their position on the chip. 

\begin{figure*}[t]
\centering
\begin{tabular}{cc}
{\includegraphics[width=6.5cm,angle=270,bb=85 100 560 720]{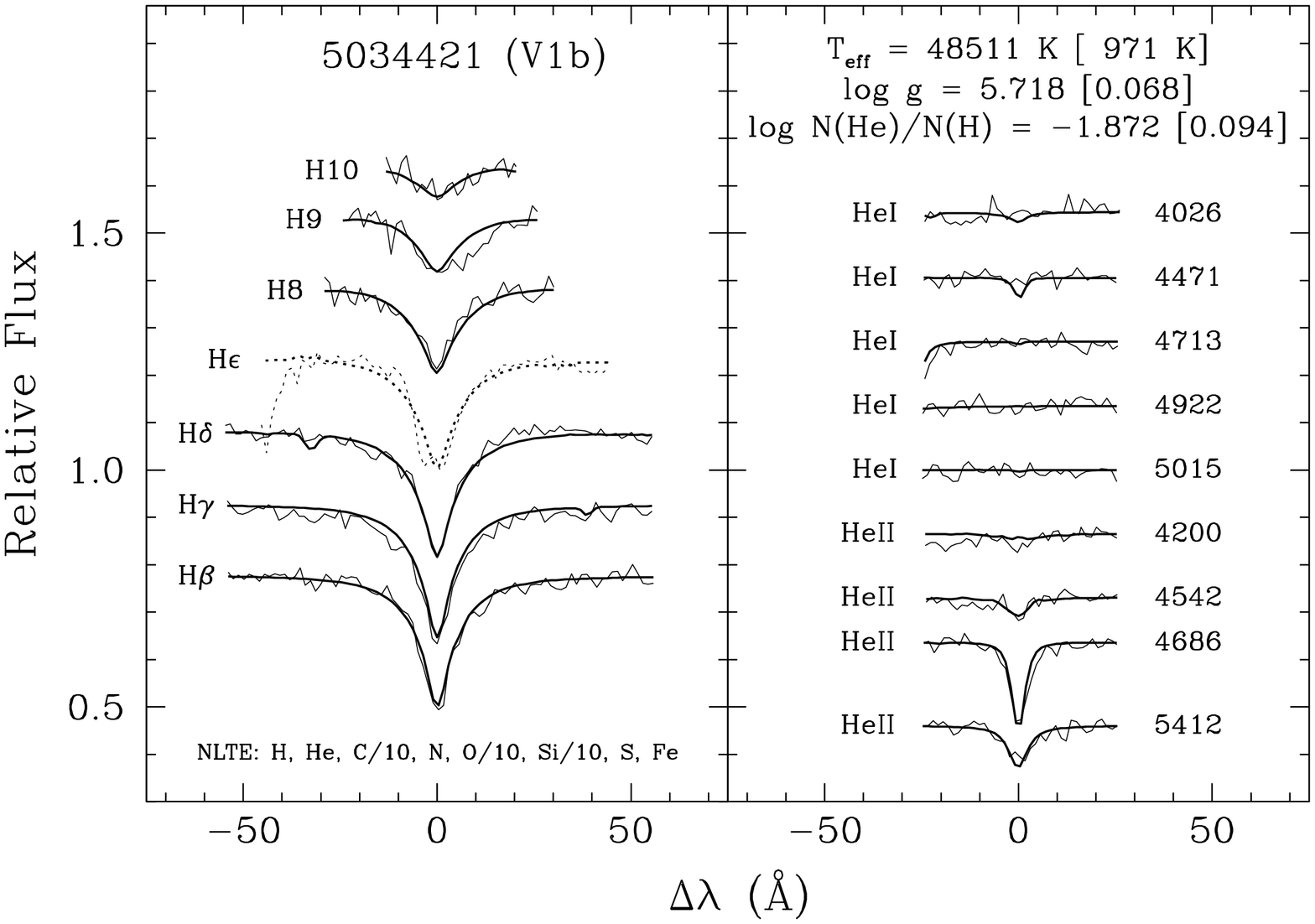}} & {\includegraphics[width=6.5cm,angle=270,bb=85 100 560 720]{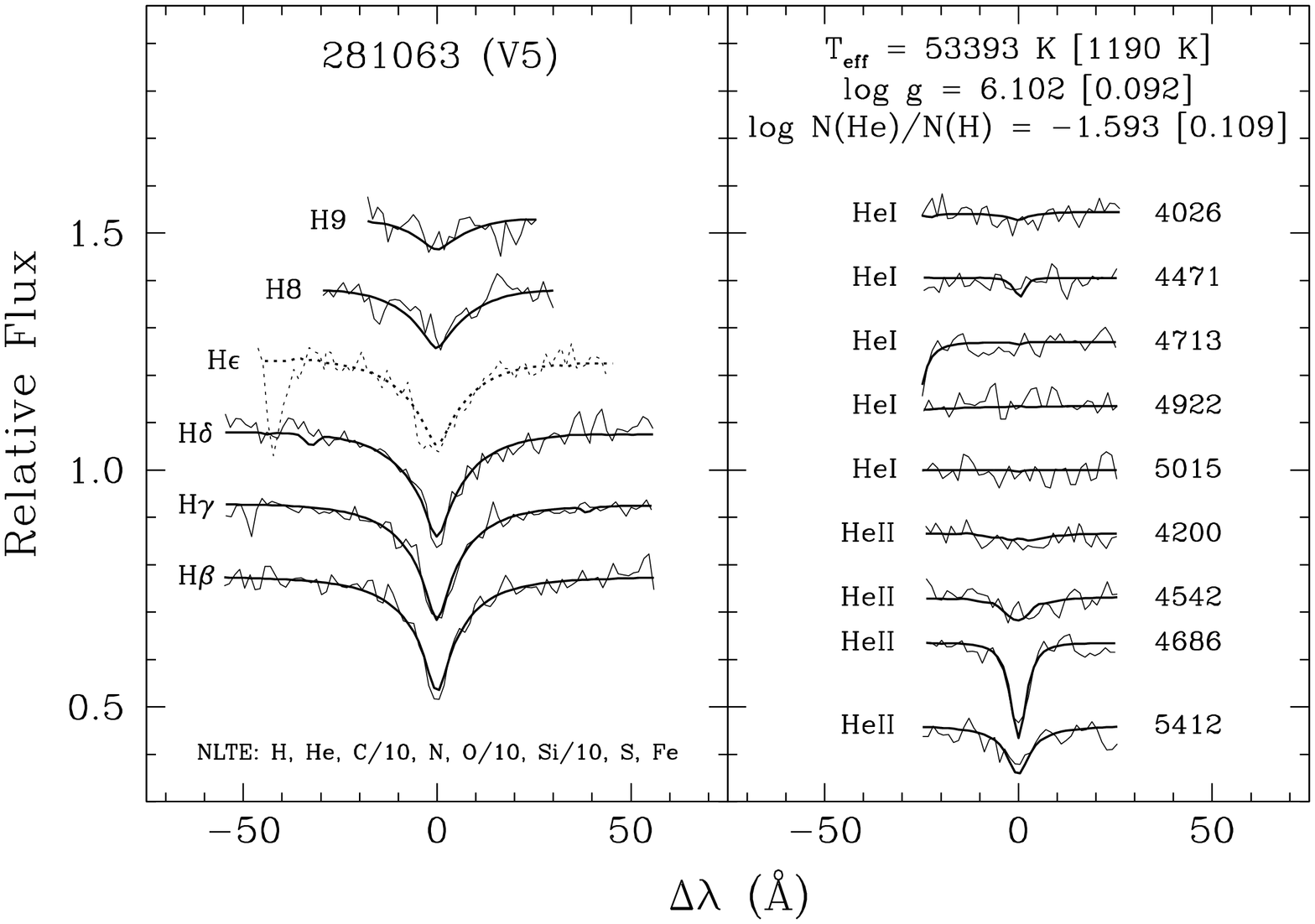}} \\
\end{tabular}
\caption{Results of the model atmosphere analysis for V1 (2013 data only) and V5. The thin lines refer to the observed spectra, while the thicker lines represent the model fit. The atmospheric parameters inferred are given, as is the composition of the models used in the fit. Note that $H{\epsilon}$ was excluded in the fitting procedure since it is contaminated by the interstellar Ca H line.
}
\label{fig:atmo}
\end{figure*}

\begin{figure*}[t]
\centering
\begin{tabular}{cc}
{\includegraphics[width=8.5cm,angle=0]{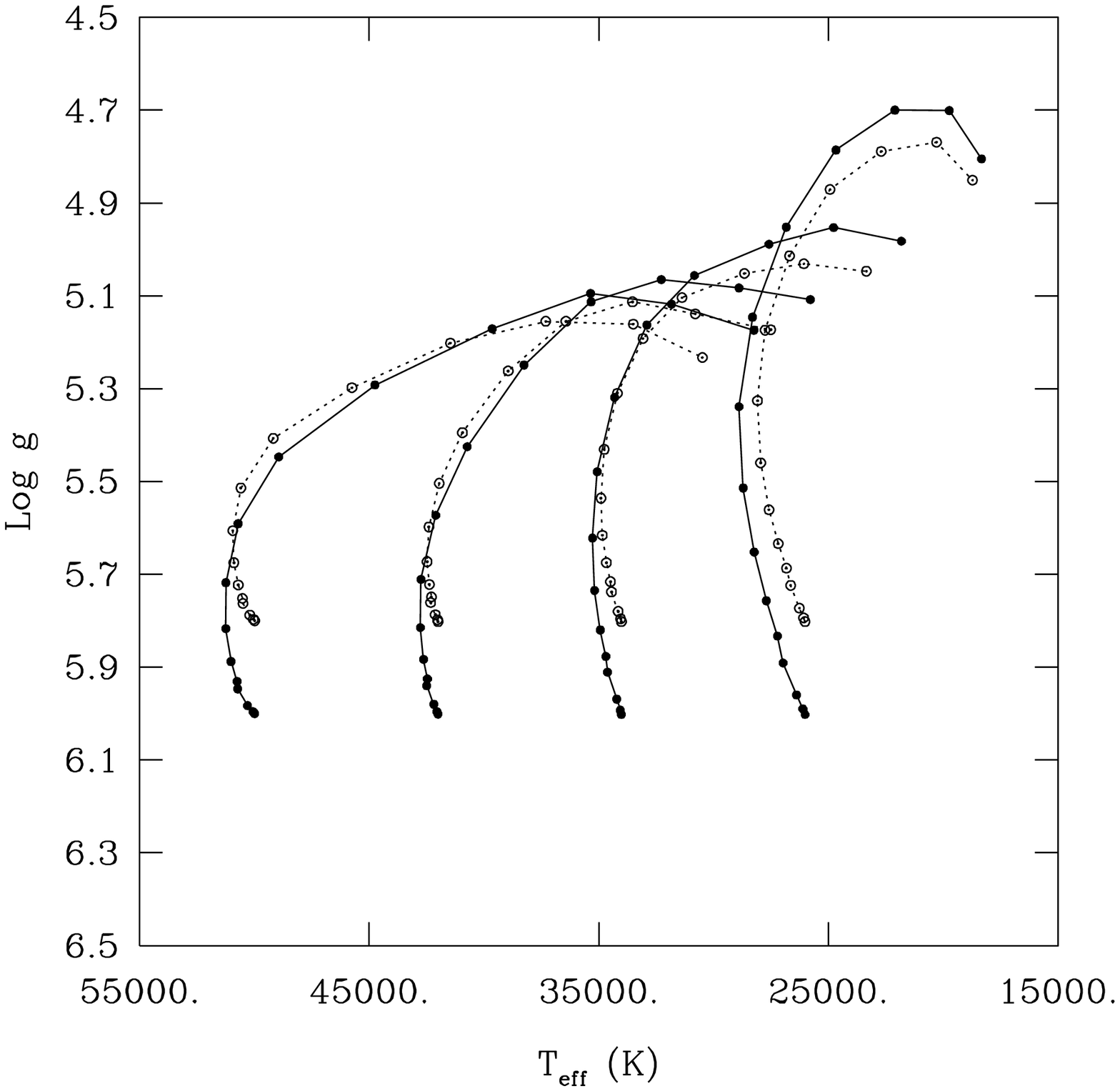}} & {\includegraphics[width=8.5cm,angle=0]{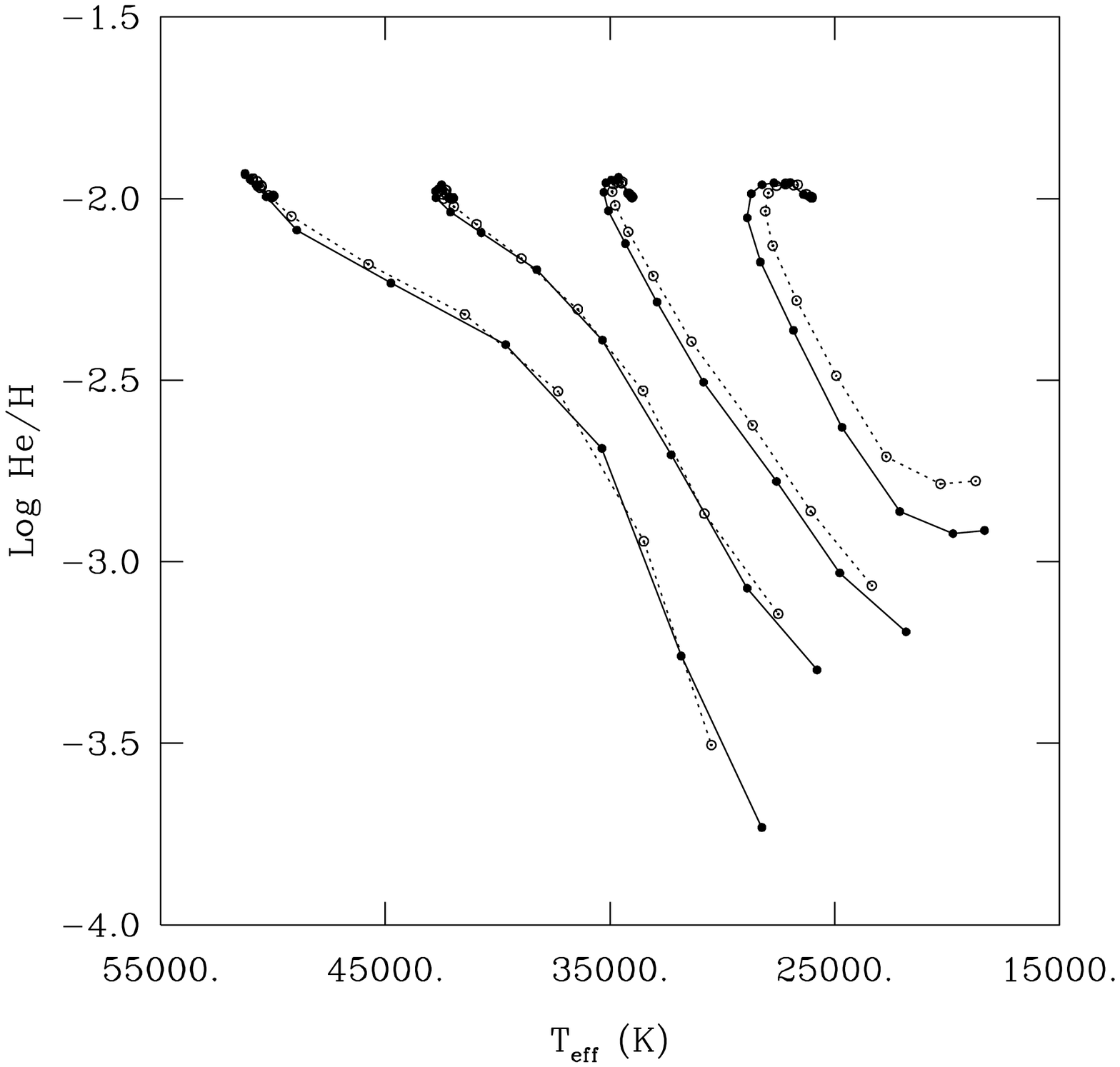}}
\end{tabular}
\caption{Results of a test designed to assess the accuracy of the atmospheric parameters inferred from hot subdwarf spectra polluted by a cooler main sequence star. The left panel shows the apparent values of $T_{\rm eff}$ and $\log{g}$ inferred from a series of polluted model spectra by fitting the same hot subdwarf spectra as used for the observational data, while the right panel highlights the effect of the pollution on the derived value of $\log{N(\rm He)/N(\rm H)}$. The tracks connect models with the same input values for the hot subdwarf component, but different levels of pollution as arising from a main sequence star (assumed to be at the same distance as the EHB star) with $T_{\rm eff}=$ 3000, 3500, 4000, 4500, 4750, 5000, 5250, 5500, 5750, 6000, 6250, 6500, 6750, 7000 and 7250 K (with $\log{g}$ = 4.5 fixed). The solid (dotted) tracks refer to a hot subdwarf component with $\log{g}$ = 6.0 (5.8) and $T_{\rm eff}$ = 50,000, 42,000, 34,000, and 26,000 K (from left to right). For the coolest main sequence component the atmospheric parameters inferred are very close to the input values for the hot subdwarf model, but they become more and more unreliable as the main sequence star becomes hotter and has a larger impact on the spectrum.  
}
\label{pollution}
\end{figure*}

\begin{table*}[t]
\caption{Atmospheric parameters inferred for our FORS2 spectroscopic sample of 47 EHB stars with clean or slightly polluted spectra. Also listed are the ACS/WFI $B$ and $R/V$ magnitudes. For the targets that overlap with our time-series photometry sample, the detection limit above which pulsations would have been detected is given. The rapid variables V1-V5 are indicated, as is the slow variable (SV).}
\centering
\begin{tabular}{lccllllccl}
\hline
\hline
{} & $T_{\rm eff}$ (K) & $\log{g}$ & $\log{(\rm He/H)}$ & Ref & $B_{ACS/WFI}$ & $R_{ACS}$ / $V_{WFI}$ & Phot. & Det. (\%) &  \\
\hline
177614 & 56,404$\pm$722 &  6.184$\pm$0.070 & $-$1.476$\pm$0.085 & ACS & 18.443$\pm$0.011 & 18.699$\pm$0.012 & efosc4 & 0.46 & \\
281063 & 53,393$\pm$1190 &  6.102$\pm$0.092 & $-$1.655$\pm$0.134 & ACS & 18.506$\pm$0.013 & 18.719$\pm$0.013 & ucam & 0.31 & V5 \\
154681 & 51,990$\pm$745 & 5.828$\pm$0.073 & $-$1.242$\pm$0.051 & WFI & 18.205$\pm$0.003 & 18.415$\pm$0.003 & efosc5 & 0.73 & V4 \\
172409$^{\ast}$ & 50,862$\pm$1127 & 5.643$\pm$0.065 & $-$2.390$\pm$0.136 & ACS &  18.151$\pm$0.015 & 18.456$\pm$ 0.016 & efosc9 & 1.08 & \\
260664$^{\ast}$ & 49,858$\pm$863 & 5.542$\pm$0.054 & $-$1.841$\pm$0.073 & ACS & 18.438$\pm$0.012 & 18.638$\pm$0.011 & ucam & 0.36 & V2 \\
177238 & 49,277$\pm$886 & 6.014$\pm$0.069 & $-$1.719$\pm$0.100 & ACS & 18.300$\pm$0.012 & 18.583$\pm$0.016 & efosc4 & 0.77 & V3 \\
5034421 & 48,500$\pm$622 & 5.798$\pm$0.046 & $-$1.819$\pm$0.060 & ACS & 18.299$\pm$0.013 & 18.547$\pm$0.012 & ucam & 0.25 & V1 \\
248955$^{\ast}$ & 48,209$\pm$1009 & 5.445$\pm$0.083 & $-$1.330$\pm$0.082 & ACS & 18.014$\pm$0.014 & 18.192$\pm$0.013 & ucam & 0.38 & SV \\
5128088$^{\ast}$ & 47,748$\pm$893 & 5.993$\pm$0.093 & $-$1.197$\pm$0.077 & WFI & 18.816$\pm$0.003 & 18.985$\pm$0.003 & ucam & 0.53 & 135227$^{\ast\ast}$ \\
5143191$^{\ast}$ & 47,552$\pm$984 & 5.475$\pm$0.082 & $-$1.388$\pm$0.081 & ACS & 18.137$\pm$0.021 & 18.339$\pm$0.043 & ucam & 0.63 & \\
5067230$^{\ast}$ & 45,351$\pm$688 & 5.740$\pm$0.107 & $-$0.539$\pm$0.074 & ACS & 18.692$\pm$0.010 & 18.892$\pm$0.011 & ucam & 1.29 & \\
165237 & 44,456$\pm$377 & 6.178$\pm$0.110 & 1.017$\pm$0.158 & ACS & 18.335$\pm$0.011 & 18.546$\pm$0.011 & no & & \\
5242616 & 43,601$\pm$626 & 5.862$\pm$0.076 & $-$1.502$\pm$0.074 & WFI & 18.427$\pm$0.004 & 18.551$\pm$0.005 & no & & \\
5039935 & 39,815$\pm$464 & 6.085$\pm$0.107 & 0.556$\pm$0.076 & ACS & 19.160$\pm$0.023 & 19.314$\pm$0.019 & ucam & 0.65 & \\
5138707 & 39,359$\pm$381 & 6.110$\pm$0.089 & 0.617$\pm$0.059 & ACS & 19.094$\pm$0.013 & 19.293$\pm$0.014 & no & & \\
5085696 & 38,757$\pm$364 & 5.711$\pm$0.084 & $-$0.102$\pm$0.053 & WFI & 18.747$\pm$0.003 & 18.744$\pm$0.004 & no & & \\
5047695 & 38,724$\pm$545 & 5.819$\pm$0.119 & $-$0.237$\pm$0.077 & ACS & 18.968$\pm$0.032 & 19.122$\pm$0.026 & ucam & 0.37 & \\
5124244 & 38,248$\pm$506 & 6.011$\pm$0.094 & $-$0.041$\pm$0.056 & WFI & 18.892$\pm$0.004 & 19.033$\pm$0.004 & ucam & 0.43 & \\
5170422 & 38,183$\pm$321 & 5.775$\pm$0.061 & $-$0.773$\pm$0.044 & WFI & 18.352$\pm$0.003 & 18.454$\pm$0.004 & ucam & 0.31 & \\
177711 & 36,984$\pm$422 & 5.783$\pm$0.073 & $-$0.530$\pm$0.052 & ACS & 18.591$\pm$0.014 & 18.796$\pm$0.014 & efosc4 & 0.48 & \\
264057 & 36,645$\pm$392 & 5.833$\pm$0.069 & $-$0.885$\pm$0.053 & ACS & 18.583$\pm$0.016 & 18.777$\pm$0.013 & ucam & 0.58 & \\
5059328$^{\ast}$ & 36,506$\pm$544 & 5.647$\pm$0.099 & $-$0.654$\pm$0.072 & ACS & 18.874$\pm$0.024 & 19.075$\pm$0.019 & ucam & 1.3 & \\
5142638 & 36,481$\pm$401 & 5.791$\pm$0.067 & $-$0.477$\pm$0.047 & WFI & 18.811$\pm$0.003 & 18.919$\pm$0.003 & efosc7 & 3.92 & \\
5220684 & 36,442$\pm$332 & 5.932$\pm$0.057 & $-$0.909$\pm$0.044 & ACS & 18.500$\pm$0.014 & 18.727$\pm$0.013 & efosc4 & 0.69 & \\
5062474 & 36,438$\pm$888 & 5.904$\pm$0.139 & $-$0.116$\pm$0.085 & ACS & 18.919$\pm$0.021 & 18.938$\pm$0.040 & ucam & 1.2 & \\
5102280 & 36,395$\pm$317 & 5.717$\pm$0.056 & $-$0.949$\pm$0.043 & WFI & 18.357$\pm$0.002 & 18.441$\pm$0.003 & no & & \\
165943 & 36,261$\pm$344 & 5.847$\pm$0.060 & $-$0.745$\pm$0.043 & ACS & 18.513$\pm$0.015 & 18.764$\pm$0.015 & no &  & \\
274052 & 36,191$\pm$486 & 5.656$\pm$0.083 & $-$0.389$\pm$0.058 & ACS & 18.720$\pm$0.014 & 18.883$\pm$0.013 & ucam & 0.36 & \\ 
5242504 & 36,072$\pm$379 &  5.807$\pm$0.066 & $-$0.532$\pm$0.048 & WFI & 18.657$\pm$0.003 & 18.824$\pm$0.004 & no & & \\
5164025 & 36,036$\pm$391 & 5.890$\pm$0.070 & $-$0.646$\pm$0.050 & WFI & 18.878$\pm$0.003 & 19.037$\pm$0.004 & ucam & 0.36 & \\
5165122 & 35,876$\pm$321 & 5.783$\pm$0.056 & $-$0.719$\pm$0.041 & WFI & 18.601$\pm$0.003 & 18.775$\pm$0.003 & ucam & 0.32 & \\
5205350 & 35,837$\pm$328 & 5.539$\pm$0.056 & $-$0.699$\pm$0.043 & WFI & 18.462$\pm$0.003 & 18.616$\pm$0.003 & no & & \\
5141232 & 35,804$\pm$356 & 5.686$\pm$0.065 & $-$0.677$\pm$0.046 & WFI & 18.578$\pm$0.003 & 18.739$\pm$0.003 & ucam & 0.35 \\
75981 & 35,122$\pm$288 & 5.789$\pm$0.049 & $-$1.107$\pm$0.041 & WFI & 18.313$\pm$0.003 & 18.513$\pm$0.003 & no & & 75981$^{\ast\ast}$ \\
53945 & 34,860$\pm$333 & 5.959$\pm$0.056 & $-$0.711$\pm$0.040 & WFI & 18.724$\pm$0.003 & 18.906$\pm$0.003 & efosc2 & 0.71 & 53945$^{\ast\ast}$ \\
5119720 & 34,775$\pm$394 & 5.836$\pm$0.066 & $-$0.861$\pm$0.050 & ACS & 18.619$\pm$0.013 & 18.778$\pm$0.013 & ucam & 0.33 & \\
5222459 & 34,668$\pm$290 & 5.793$\pm$0.048 & $-$0.722$\pm$0.036 & WFI & 18.825$\pm$0.003 & 18.952$\pm$0.003 & efosc2 & 0.81 \\
5180753 & 34,140$\pm$283 & 5.783$\pm$0.049 & $-$1.490$\pm$0.051 & ACS & 18.400$\pm$0.012 & 18.599$\pm$0.016 & efosc9 & 1.18 & \\
5142999 & 33,843$\pm$380 & 5.641$\pm$0.061 & $-$1.155$\pm$0.056 & WFI & 18.549$\pm$0.003 & 18.735$\pm$0.003 & ucam & 0.27 & 164808$^{\ast\ast}$ \\
5243164 & 31,614$\pm$245 & 5.408$\pm$0.040 & $-$2.656$\pm$0.134 & WFI & 18.206$\pm$0.003 & 18.359$\pm$0.004 & no & & 75993$^{\ast\ast}$ \\
168035 & 31,458$\pm$431 & 5.626$\pm$0.070 & $-$3.614$\pm$0.557 & ACS & 17.929$\pm$0.010 & 18.084$\pm$0.011 & no & & \\
5185548$^{\ast}$ & 31,191$\pm$338 & 5.546$\pm$0.056 & $-$4.365$\pm$1.064 & WFI & 18.116$\pm$0.002 & 18.251$\pm$0.002 & ucam & 0.24 & \\
5262593 & 30,762$\pm$255 & 5.497$\pm$0.042 & $-$3.393$\pm$0.417 & WFI & 17.810$\pm$0.002 & 17.938$\pm$0.002 & no & & \\
175411$^{\ast}$ & 29,213$\pm$293 & 5.454$\pm$0.047 & $-$3.176$\pm$0.292 & ACS & 18.698$\pm$0.016 & 18.713$\pm$0.031 & efosc9 & 1.62 & \\
204071 & 28,761$\pm$657 & 5.570$\pm$0.095 & $-$3.074$\pm$0.186 & ACS & 17.929$\pm$0.021 & 18.100$\pm$0.014 & no & & \\
5139614 & 27,894$\pm$379 & 5.521$\pm$0.058 & $-$3.758$\pm$1.185 & ACS & 18.823$\pm$0.012 & 18.934$\pm$0.010 & ucam & 0.63 & \\
5238307 & 27,108$\pm$261 & 5.512$\pm$0.039 & $-$2.287$\pm$0.073 & ACS & 18.590$\pm$0.013 & 18.673$\pm$0.018 & no & & \\
\hline
\end{tabular} \\
$^{\ast}$ spectrum polluted by cooler star. The estimate of $\log{g}$ should be seen as a lower limit only. \\
$^{\ast\ast}$ ID from \citet{moehler2011}
\label{atmo}
\end{table*}

Prior to the detailed atmospheric analysis, the spectra obtained for each of 60 distinct targets were carefully inspected. All showed a redshift of $\sim$4 \AA\, indicating that they indeed belong to $\omega$ Cen, which has a motion of $\sim$ 230 km/s relative to Earth. Two of the targets turned out to not be EHB stars, and were discarded. The remaining 58 bona fide EHB stars were then divided into two groups: those with ``clean'' and those with ``polluted'' spectra. Targets in the first group show the spectrum of an apparently single star, while those in the second group have spectra that are polluted to some degree by a cooler star. Note that this does not imply that the ``clean'' sample corresponds to single EHB stars and the ``polluted'' sample contains only binaries. Given the crowdedness of the field a lot of the EHB stars are contaminated by stars that happen to fall nearby but are not necessarily associated in any physical way, and the apparently single stars may have an unseen faint companion such as a white dwarf or a cool main sequence star. We can however rule out an early type main sequence companion (such as the F or G stars observed for many subdwarf B stars among the field population) for the ``clean'' EHB stars.

A spectrum was declared ``clean'' if the following criteria were met: 1) The slope of the continuum is normal, i.e. relatively flat from blue to red in the unnormalised, not flux calibrated spectrum. Pollution from a cooler companion necessarily leads to a steeper increase of the flux towards longer wavelengths. 2) The Ca II doublet (3933 \AA\ K line and 3968 \AA\ H line, the latter is partially fused with H$\epsilon$) is blue-shifted relative to the rest of the spectrum, implying that it arises from interstellar absorption rather than a contaminating star in the (red-shifted) cluster. The Ca doublet is not seen in the restframe of the spectrum. 3) The G-band (made up of the feature due to Ca I at 4226 \AA, the CH molecular band at 4300 \AA\ and the Fraunhofer G line at 4307 \AA) is not detected. 4) The Mg I triplet (5167, 5172, 5183 \AA) is not detected. 5) The Na I doublet (5890 and 5896 \AA) is blue-shifted compared to the rest of the spectrum and thus arises from interstellar absorption. Conversely, the presence of the G-band and Mg I triplet as well as Ca II and Na I in the restframe of the spectrum was interpreted as significant pollution by a cooler star. The difference between some of the more polluted and some ``clean'' spectra obtained is nicely illustrated in Fig. \ref{gpol}.

Out of the 58 EHB spectra obtained, 38 were classified as ``clean'' and 20 as ``polluted'' by a cooler star to some degree. We attempted to fit the Balmer and He lines using our bank of non-LTE synthetic spectra \citep[see, e.g.][for details]{latour2011}. These are based on model atmospheres including metals with the composition derived by \citet{blanchette2008} for hot subdwarfs on the basis of FUSE spectra. Two example fits (for V1 and V5) are shown in Fig. \ref{fig:atmo} for illustrative purposes. Note that the model atmosphere fits and parameters inferred for V1--V4 are already published in \citet{randall2011} based on an initial analysis of the March 2011 spectroscopy only. V1 and V2 were re-observed in 2013 and the values presented in Table \ref{atmo} are based on an average of the two observations.

In order to assess the reliability of the atmospheric parameters derived for the ``polluted'' sample we carried out some tests using model spectra of hot subdwarfs to which we added the relative contributions of main sequence companions of varying effective temperatures. These composite model spectra were then fit using the same models and procedure as applied to the real, observed spectra, and apparent values of $\log{g}$, $T_{\rm eff}$ and $\log{q(\rm He)}$ were derived. The results of this are illustrated in Fig. \ref{pollution}. It can be seen that spectra suffering from slight contamination will yield quite reliable helium abundances and effective temperatures (although these will be overestimated a little), while the surface gravity will be underestimated due to the deepening of the Balmer lines from the lower-gravity star. Spectra that are strongly contaminated on the other hand will produce completely unreliable atmospheric parameters. The only way to analyse these properly would be to use an iterative modelling approach that fits both the subdwarf and the main sequence component, however this would require spectra of higher quality and more accurate flux calibration than those in our sample.

Out of the 20 ``polluted'' targets, 9 have inferred apparent values of $\log{g}\gtrsim$ 5.4 and show only slight to moderate pollution in their spectra. Taking Fig. \ref{pollution} as a guideline we then expect the values inferred for $T_{\rm eff}$ and $\log{N(\rm He)/N(\rm H)}$ to provide a good indication of the actual values, while $\log{g}$ is most likely underestimated and should be interpreted as a lower limit. We include these 9 targets in the results presented below, while the remaining 11 strongly polluted targets are excluded from further analysis. Our spectroscopic sample then contains 47 EHB stars for which useful atmospheric parameters could be inferred. Note that 3 targets (53945, 75981 and 5142999) overlap with the spectroscopic sample of hot horizontal branch stars in $\omega$ Cen for which \citet{moehler2011} derived atmospheric parameters; our estimates are consistent within the errors. Another two of our targets were also observed by \citet{moehler2011}, but no atmospheric parameters were derived.

\section{The EHB instability strip in $\omega$ Cen}

\subsection{The empirical instability strip}

\begin{figure}[t]
\centering
\includegraphics[width=9.0cm,bb=70 150 550 700,angle=0]{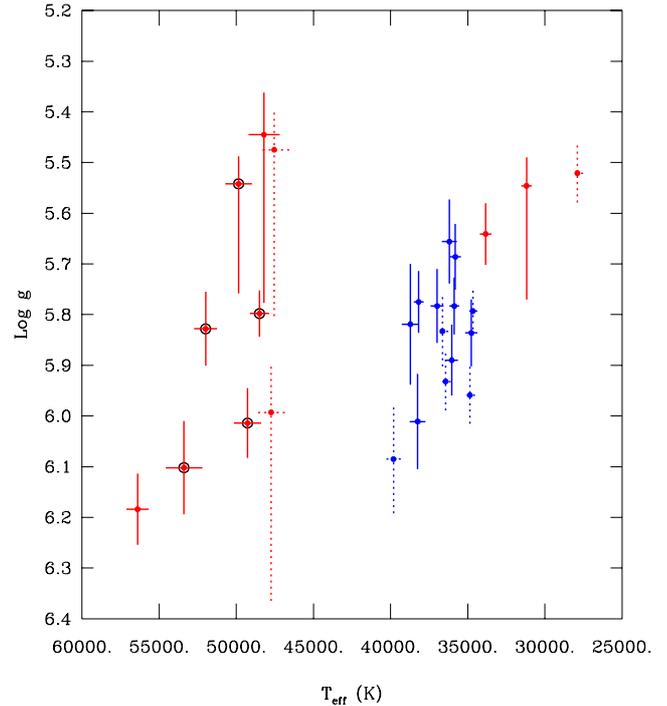}
\caption{Observed EHB star instability strip in $\omega$ Cen, based on the 26 targets listed in Table \ref{atmo} for which the achieved detection threshold for pulsations is less than 1\%. The five rapid pulsators are marked by large open circles. Red symbols indicate H-rich EHB stars, blue ones show He-rich targets (compared to Solar). The error bars refer to the formal uncertainties on the atmospheric parameters and have been extended somewhat arbitrarily towards higher surface gravities for the spectra classified as polluted since the derived values of $\log{g}$ only constitute a lower limit for these stars. Dotted error bars indicate targets where the detection threshold of the photometry lies above 0.5\% (but below 1.0\%) and low-amplitude pulsations cannot be excluded.}    
\label{strip}
\end{figure}

In order to define the observed EHB instability strip in $\omega$ Cen we combined the results from our EFOSC2/ULTRACAM photometry and the FORS2 spectroscopic targets, focusing only on objects for which both a meaningful detection limit for pulsations as well as reliable atmospheric parameters are available. These can be identified easily in Table \ref{atmo}. If a spectropscopic target was also observed in the photometry, we list the field as well as the detection threshold, defined as the lower of 3.7 times the mean noise level or the highest peak. On the basis of these data we compute the empirical EHB star instability strip shown in Fig. \ref{strip}. Looking at the plot, the five rapid variables appear to cluster together and form a relatively well-defined instability strip between about 48,000 and 54,000 K. From the current sample it is not clear whether this instability strip is pure, i.e. whether all $\omega$ Cen EHB stars within this temperature range pulsate, but there is no convincing evidence that this is the case. Remember that our spectroscopic sample is strongly biased towards the pulsators, as they were specifically selected as spectroscopic targets. Therefore, the fact that we find no non-pulsating EHB stars that lie clearly within the region where pulsators are found is not statistically significant at this stage. Similarly, the boundaries of the empirical instability region are somewhat fuzzy due to the relatively small number of targets for which we have both reliable spectroscopy and time-series photometry.  

The other thing to note from Fig. \ref{strip} is that we find no rapid pulsators at lower temperatures, particularly in the sdBV$_r$ instability strip between $\sim$29,000--36,000 K. However, looking more closely we find that we observed only 2 stars that have the appropriate atmospheric parameters for sdBV$_r$ stars (note that those are all H-rich) to a photometric accuracy where we can exclude pulsations typical for this class of variable. Since the sdBV$_r$ instability strip in the field is far from pure this means we cannot exclude the existence of these objects in $\omega$ Cen from the available data. Certainly, sdBV$_r$ candidates (i.e. H-rich sdBs around 33,000 K) appear to be far less common in $\omega$ Cen than in the field relative to the remaining hot subdwarf population \citep[see][]{latour2014}. Combined with the inherent observational difficulties when searching for small amplitude variations in faint stars in very crowded fields this means that any sdBV$_r$ stars in $\omega$ Cen (or other globular clusters) will be very challenging indeed to find.

\subsection{The theoretical instability strip}

We extended our Montr\'eal ``second-generation'' models that are routinely used for the interpretation of sdBV$_r$ stars \citep[see, e.g.][]{brassard2001} to higher temperatures in order to investigate the presence of unstable $p$-modes also for sdO stars. The new extended grid covers the 20,000--78,000 K temperature range and surface gravities between $\log{g}$=4.8 and 6.4. The other two free parameters in the models are the total stellar mass $M_{\ast}$ and the logarithmic depth of the transition zone between the He-rich core and the H-dominated envelope, but since they do not impact the qualitative instability calculations presented here we kept them fixed at representative values of $M_{\ast}$=0.48 $M_{\odot}$ and $\log{q(\rm H)}$=$-$4.0. Recall that the ``second-generation'' models incorporate traces of Fe that are levitating in a pure H background under the assumption that an equilibrium has been reached between radiative levitation and gravitational settling. It is the opacity bump associated with a local overabundance of iron in the driving region that allows pulsations to be driven via the $\kappa$-mechanism \citep{charp1996,charp1997}.    

\begin{figure}
\centering
\includegraphics[width=9.0cm,bb=70 120 550 670,angle=0]{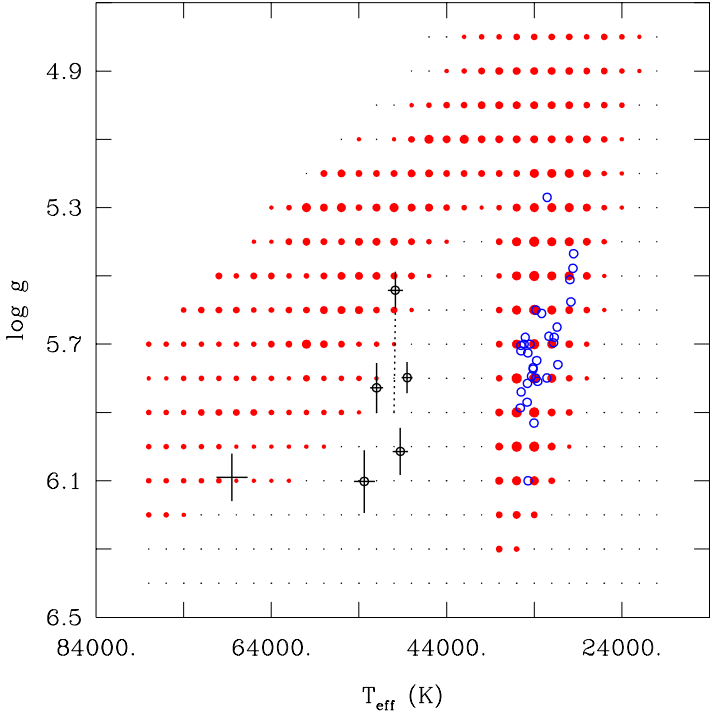}
\caption{Theoretical instability strip for rapid $p$-mode pulsations in hot subdwarfs. The small black dots identify grid points corresponding to stable models. Each red point identifies a model where unstable modes are predicted, the size of the dot being proportional to the number of excited modes. The large blue dots show the location of sdBV$_r$ variables among the field population, while the black cross indicates the one known rapid sdO variable in the field. The five sdO variables we found in $\omega$ Cen are represented by black circles, the dotted extension to higher $\log{g}$ for V2 indicating that the polluted spectrum obtained yields only a lower limit on the surface gravity.}  
\label{jaws}
\end{figure}

\begin{figure*}[h]
\centering
\begin{tabular}{cc}
{\includegraphics[width=8.5cm,angle=0, bb=70 70 550 720]{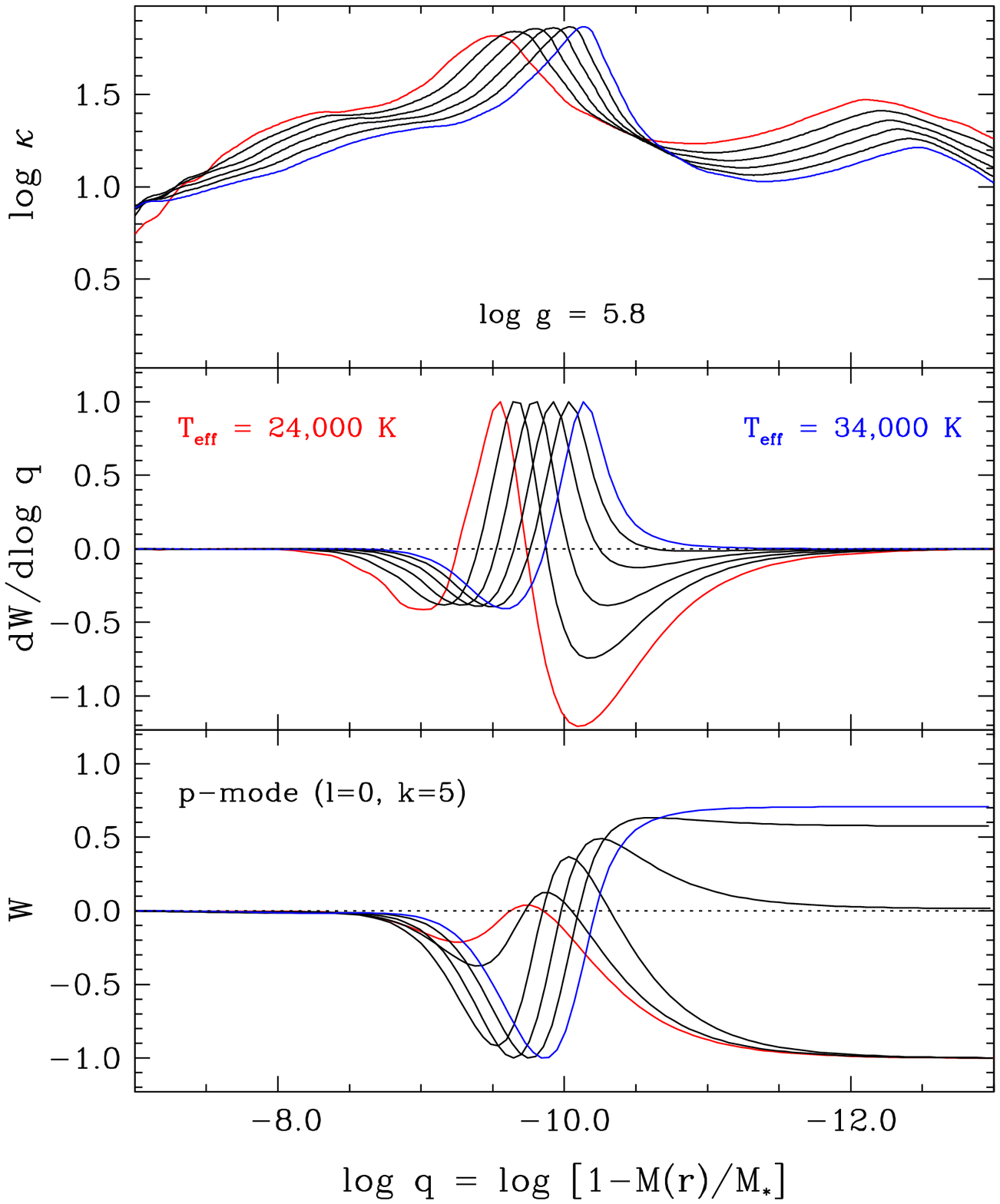}} & {\includegraphics[width=8.5cm,angle=0,bb=70 70 550 720]{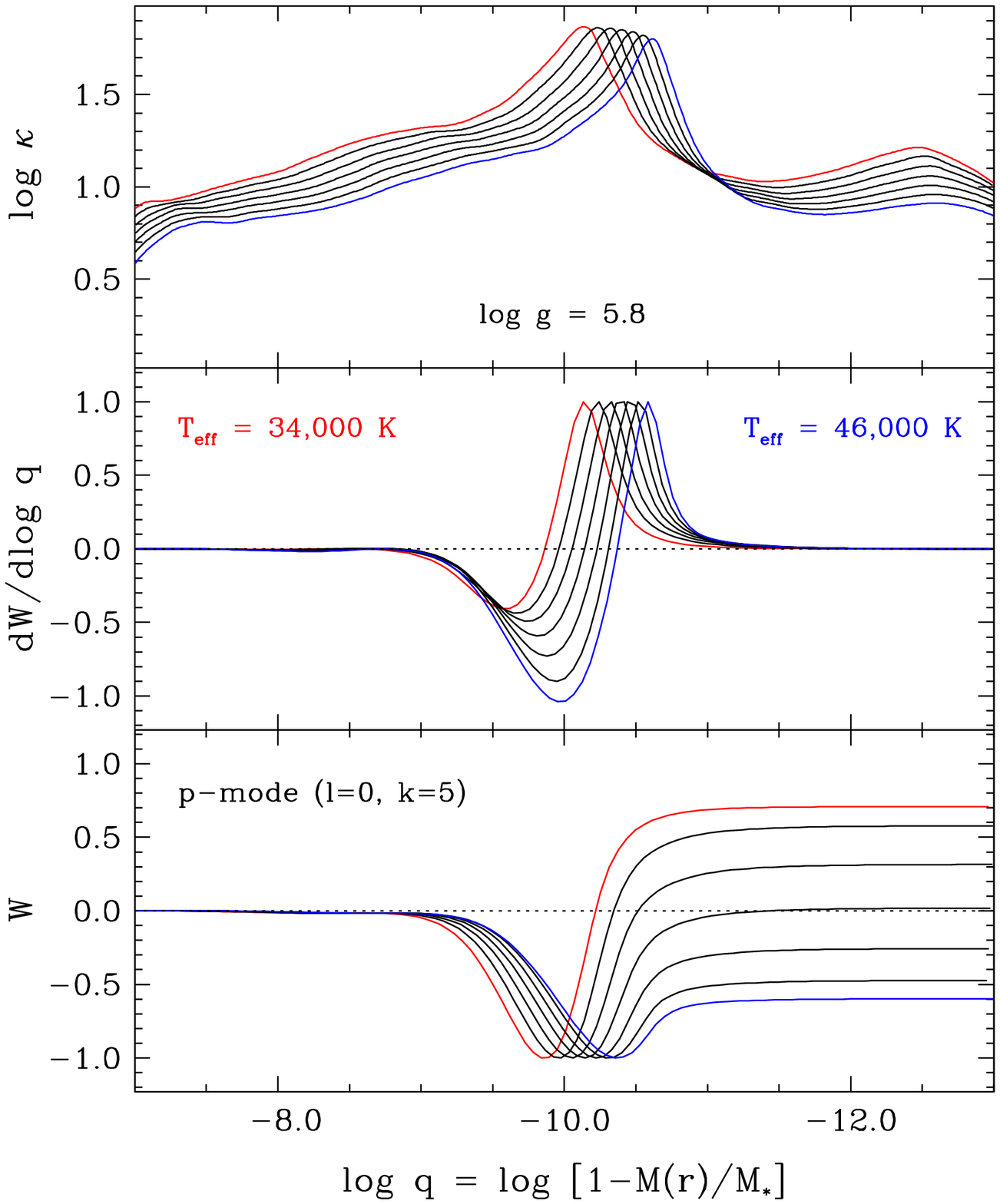}} \\
{\includegraphics[width=8.5cm,angle=0, bb=70 70 550 720]{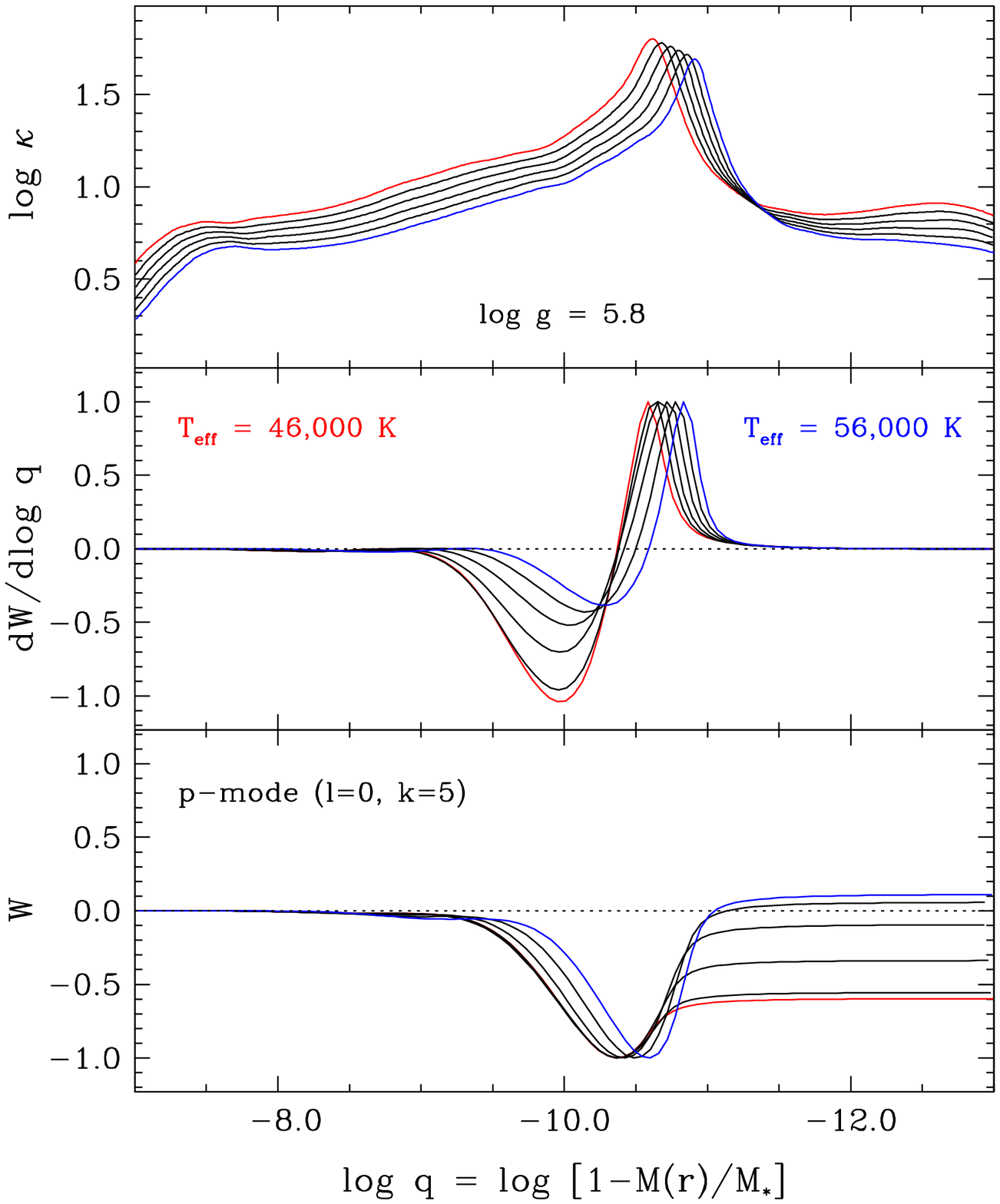}} & {\includegraphics[width=8.5cm,angle=0,bb=70 70 550 720]{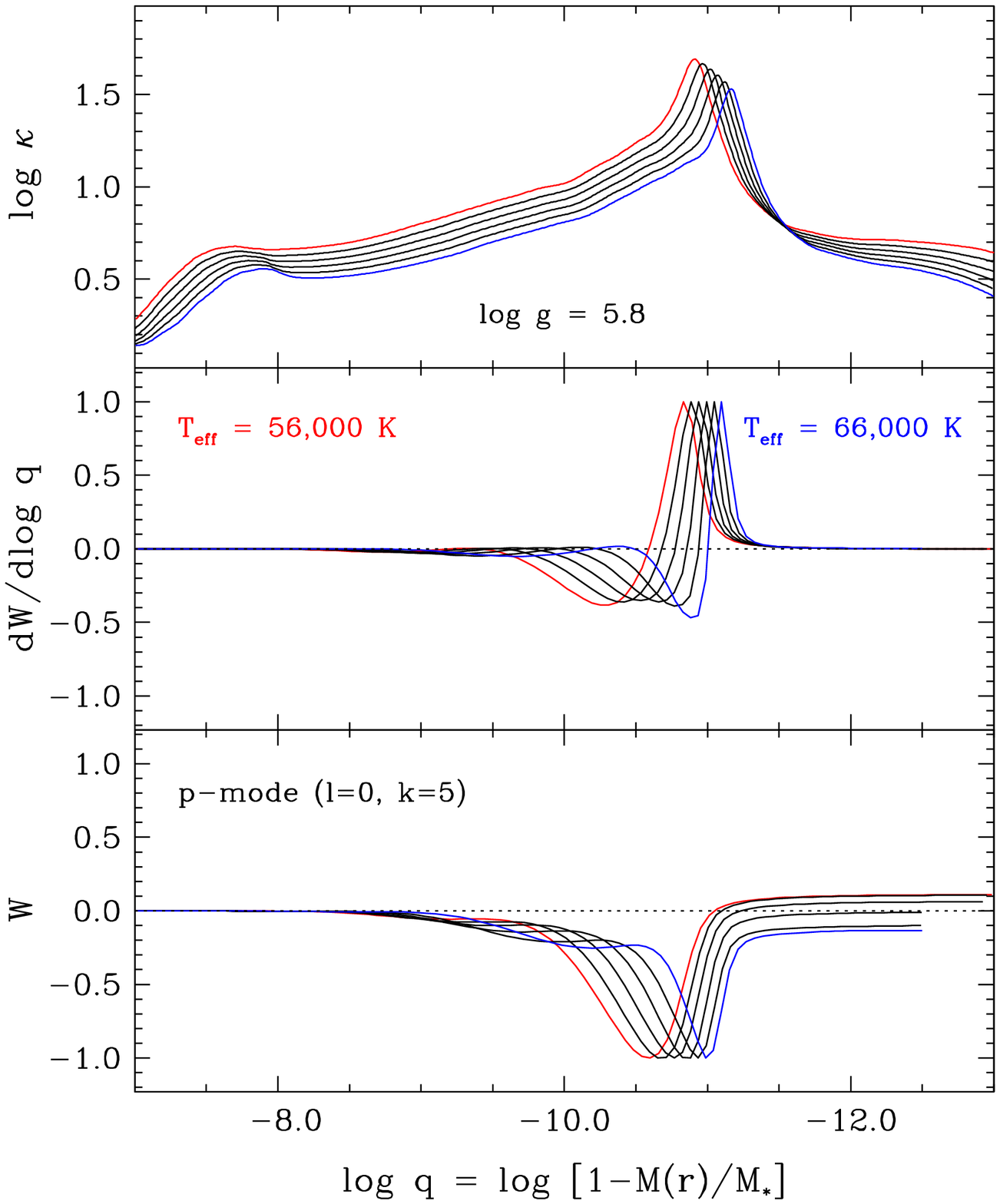}}
\end{tabular}
\caption{Sequence of models where the effective temperature is varied from 24,000 K to 66,000 K in steps of 2000 K over the four plots as indicated. The top panel of each plot shows the logarithm of the Rosseland mean opacity, the middle panel shows the derivative of the work integral and the bottom panel shows the cumulative work integral as a function of stellar depth for a representative $p$-mode (where the centre of the star is to the left and the surface to the right). If the latter is positive at the surface, the mode is excited, otherwise it is stable.}
\label{kappa}
\end{figure*}

In the instability calculations presented here we focus on low-order radial modes with $P\sim$15--1200 s, but the non-adiabatic results obtained are representative also of higher degree modes. Fig. \ref{jaws} shows the results of our analysis. It can be seen that in addition to the well-known sdBV$_r$ instability strip around 34,000 K there is a second region of instability at similar surface gravities but much higher effective temperatures. This was already suspected from an exploratory study of pulsation driving around the field sdO V499 Ser \citep{fontaine2008}, but never investigated in a systematic way. The existence of a bifurcated $p$-mode instability strip can be understood in terms of the way the iron abundance profile and hence the resulting opacity profile evolve as a function of the effective temperature of the model. This is illustrated in Fig. \ref{kappa}, where we show the profiles of the logarithmic Rosseland mean opacity $\log{\kappa}$, the cumulative work integral of a representative mode $W$, and its derivative as a function of stellar depth $dW / d\log{q}$ for a sequence of representative models with fixed log$g$ = 5.8 and $T_{\rm eff}$ increasing in steps of 2000 K from 24,000 K (top left) to 66,000 K (bottom right) as indicated. Note that the depth within the star is expressed in terms of a logarithm to better visualise the outer layers of the star, in particular the envelope where the interesting "action" occurs. Looking at the top panels, we can see how the iron opacity bump (at $\log{q(\rm H)}\sim$ $-$9.5 for the 24,000 K model) is pushed towards the stellar surface as the temperature of the model is increased. This directly affects the work function of the pulsation modes that may potentially be excited in the envelope of the model. Here, we show two instructive quantities for a representative radial mode with radial order $k$=5 \citep[for more details on the quantities and notation used please see][]{charp2001}. The derivative of the work integral gives the net amount of energy locally gained (or lost) by the material displaced by the mode during one cycle of pulsation: where it is positive, the corresponding region contributes to destabilising or exciting the mode, where it is negative the mode experiences a stabilising or damping effect. It then follows that modes for which the cumulative work integral (summed from the centre towards the surface) is positive are excited, while those where it is negative are stable. Following the model sequence, we see that at 24,000 K the damping dominates and the mode is stable. With increasing temperature, the damping becomes less important and between 30,000--36,000 K the mode becomes unstable in the sdBV$_r$ instability strip. Then the damping again becomes stronger than the excitation before diminishing once again and allowing the mode to be driven between 54,000 K and 58,000 K.  Note however that the driving in this hotter instability strip is rather weak in relative terms compared to the sdBV$_r$ region. 

Comparing the predicted instability regions in $T_{\rm eff}-\log{g}$ space to the location of the real pulsators in Fig. \ref{jaws} we see that while the known sdBV$_r$ stars match the models perfectly, the $\omega$ Cen pulsators are cooler than the red edge of the hotter instability region. This discrepancy may be partially explained by the spectroscopic temperatures likely being underestimated for these hot stars. Indeed, it is thought that analyses based on optical spectroscopy yield systematically cooler temperatures for stars hotter than $\sim$50,000 K due to the so-called Balmer-line problem \citep{napi1993}. For the few cases where both high-quality optical and UV data are available and have been exploited for very hot compact evolved stars, the (more reliable) UV spectroscopy yielded temperature estimates over 10,000 K higher than the optical spectra \citep[see][for a recent example]{latour2015}. We plan to refine the temperatures of the $\omega$ Cen variables on the basis of HST-COS spectra in the future, however this is outside of the scope of this study and will be presented in a future publication. For the time being, it is enough to keep in mind the possibility that the temperatures derived for the hotter stars of our sample may be systematically underestimated.

\begin{figure}[t]
\centering
\includegraphics[width=9.0cm,bb=20 150 550 700,angle=0]{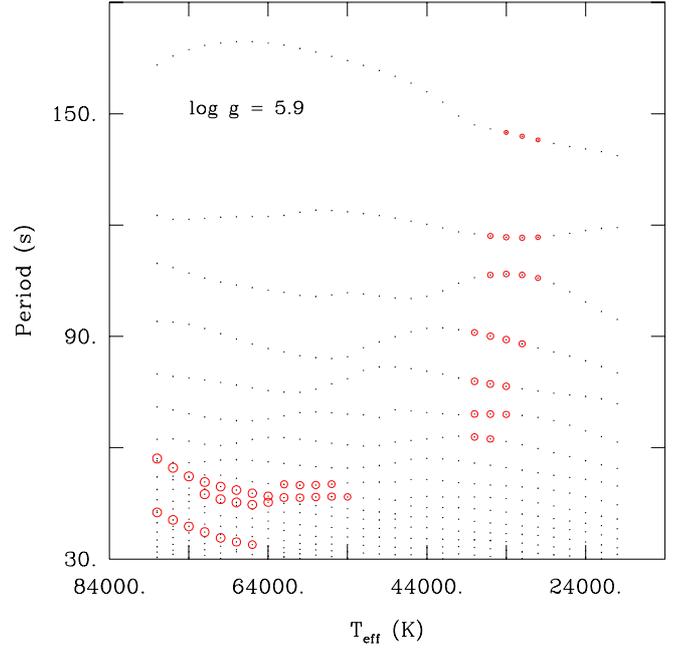}
\caption{Periods excited as a function of temperature for a sequence of representative sdB/sdO models. Each red circle refers to an excited mode, while black dots indicate stable modes. 
}  
\label{periods}
\end{figure}  

However, it is clear that simply revising the spectroscopic temperatures of the pulsators will not solve the problem and bring their observed pulsational properties in line with what is predicted from our models. This is apparent from Fig. \ref{periods}, where we show the periods excited for a series of sdB/O star models. While the periodicities predicted for sdBV$_r$ stars around 34,000 K nicely overlap with the $\sim$100--150 s periods typically observed for these stars, the modes excited at higher temperatures have very short periods below 60 s, significantly shorter than the 85--125 s pulsations observed for the $\omega$ Cen variables. We are well aware of the limitations and missing ingredients in the ``second-generation'' models, notably the fact that the opacity profile takes into account only radiatively levitating iron, while neglecting other important contributors. Nickel in particular is thought to substantially affect the opacity profile \citep[see][]{jeffery2006,hu2011}, and indeed may boost driving at temperatures around 50,000--70,000 K. Also, processes other than diffusion such as a stellar wind or weak turbulent mixing are not taken into account. Our models clearly need to be refined in order to accurately reproduce the pulsations observed in the $\omega$ Cen pulsators, and this is an ongoing long-term project. However, since the mismatch between the observed and predicted pulsations is at the quantitative rather than the qualitative level we do believe we have identified the driving mechanism for these hotter sdO pulsators to be the same basic $\kappa$-mechanism responsible for the other classes of hot subdwarf pulsators.  

\subsection{Is the $\omega$ Cen instability strip unique?}

The discovery of rapidly pulsating H-rich sdO stars in $\omega$ Cen was completely unexpected, since such objects were neither known nor predicted from models. But of course, this is the beauty of observational astronomy: looking for one thing often leads to finding something completely different. Following the discovery of the pulsators in $\omega$ Cen, a systematic search was launched to identify counterparts among the field population by \citet{johnson2014}. Out of their sample of 36 field sdO stars, 6 are H-rich sdOs that fall within the empirical $\sim$48,000--54,000 K $\omega$ Cen instability strip. None of their targets show periodic oscillations down to a (very stringent) detection threshold of 0.08\%. While it is possible that we are limited by small number statistics here, it does seem indicative that out of $\sim$150 EHB targets monitored photometrically in $\omega$ Cen we found 5 H-sdO pulsators, whereas none were detected among the 600+ observed field hot subdwarfs observed to date with much higher accuracy. Unlike the longer period oscillations in the sdBV$_s$ stars, the short and high-amplitude periodicities of the $\omega$ Cen variables would not have been overlooked in noisier datasets from smaller telescopes. On the other hand, given the limited number of field sdO stars monitored and the possibility that the pulsator fraction among the H-sdOs in the instability strip around 50,000 K is only say 10\%, it is possible that such objects are simply extremely rare among the field population and still remain to be found. 

The lone known field sdO pulsator V499 Ser appears to be quite distinct from the $\omega$ Cen pulsators in that it is around 20,000 K hotter and shows a mild He enhancement ($\log{N(\rm He)/N(\rm H)}\sim -$0.64), while all the $\omega$ Cen variables are extremely He-poor. On the other hand, the pulsational properties are very similar to those reported here: V499 Ser shows a very dominant peak at 119.3 s with an amplitude of 3.5\% and then several much lower amplitude ($<$0.5\%) peaks with periods between 60 and 120 s \citep{woudt2006}. And indeed, the hotter part of the instability strip predicted from Fig. \ref{jaws} does cover a wide temperature range that could potentially encompass both the $\omega$ Cen variables and V499 Ser. More sophisticated models, and also more observed pulsators are needed in order to settle the relationship between the different sdO oscillators.  

An additional complicating factor in our understanding of different classes of EHB pulsators is the recent discovery of 6 rapid pulsators in the core of NGC 2808 based on HST observations by \citet{brown2013}. Out of a sample of $\sim$100 hot evolved stars, 6 showed rapid luminosity variations with (FUV) amplitudes of $\sim$ 2--7\% and periods between 85 and 149 s. Three of the pulsators were also observed spectroscopically with STIS, allowing the effective temperature and atmospheric helium abundance to be constrained if not accurately determined due to the relatively high noise in the data. Intriguingly, the NGC 2808 pulsators seem a rather inhomogeneous bunch, and none of them show a similar combination of atmospheric and pulsational properties as the $\omega$ Cen variables, or any of the EHB variables known in the field for that matter. Their existence is puzzling to say the least, and follow-up observations allowing a better determination of their atmospheric parameters are urgently needed. Unfortunately, their location in the centre of NGC 2808 precludes ground-based observations, leaving HST as the only option. 

\begin{figure}[t]
\centering
\includegraphics[width=9.0cm,bb=20 150 550 700,angle=0]{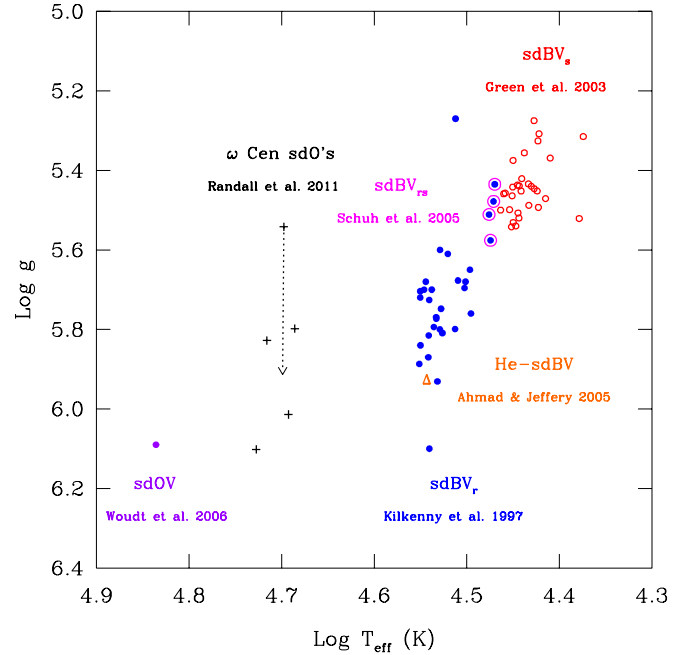}
\caption{An overview of the different types of EHB pulsator currently known among the field population together with the $\omega$ Cen pulsators found during our survey. The discovery paper is indicated for each class of variable.   
}  
\label{ehb_pulsators}
\end{figure} 

\section{Conclusion}

Our observational time-series photometry and spectroscopy survey of EHB stars in $\omega$ Cen uncovered a new class of H-rich sdO pulsators with short periods between 84 and 124 s and effective temperatures around 50,000 K. In total, we discovered five such pulsators, four of which show high-amplitude pulsations above $\sim$ 2\%, and one of which exhibits much lower-amplitude luminosity variations around 0.5\%. The overall yield of pulsators in our sample is around 3\%, slightly higher if we consider only targets for which we obtained high-quality light curves. From our observations it is not clear whether the $\omega$ Cen instability strip is pure, and it seems quite feasible that pulsators and non-pulsators co-exist with similar atmospheric parameters. We detected no representatives of the well-known field sdBV$_r$ stars in $\omega$ Cen, but from our sample cannot exclude that they exist at a similar pulsator to non-pulsator fraction as those in the field. Quite interestingly, no direct counterparts to the $\omega$ Cen pulsators have been found in the field or in other globular clusters, despite dedicated surveys. Fig. \ref{ehb_pulsators} gives a nice overview of the location of the different classes of known EHB pulsator among the field population compared to those found in $\omega$ Cen in $\log{g}-T_{\rm eff}$ space. 

In addition to the rapid pulsators, our photometry revealed one target that shows a slower luminosity variation with a period of $\sim$2700 s. The most intriguing explanation for this is an ellipsoidal deformation of the hot subdwarf in a close binary with a massive companion, however other possibilities cannot be excluded at this point. Radial velocity measurements are needed in order to draw any more definite conclusions here.

We believe we have identified the driving mechanism for the rapid pulsations in the $\omega$ Cen variables to be the same $\kappa$-mechanism that is responsible for the excitation of the oscillations in both the short- and long-period variable sdB stars in the field. Indeed, our non-adiabatic calculations show that the well-known $p$-mode instability strip encompassing the sdBV$_r$ pulsators is complemented by another instability region of shorter periods at higher temperatures. At the quantitative level however there are some discrepancies between the predicted and the observed pulsations: the models excite pulsations at higher temperatures and shorter periods than what is observed. We plan to follow up on this in the future with FUV spectroscopy and more sophisticated models. The former should allow us to reliably pin down the effective temperatures for these very hot stars, which are likely underestimated from the optical spectroscopy presently available. On the modelling side, we need to include other elements besides iron that affect the opacity profile in the driving region, and also take into account the variable H-He composition of the envelope from one star to the next. 

The more we observe EHB stars in different environments, the more complicated the picture gets. In addition to the atmospheric parameters and binary fraction also the pulsational properties of EHB stars appear to markedly differ between the field and globular cluster population, and even from one globular cluster to the next. This must point to distinct evolutionary channels dominating the formation of these stars in the different environments. However, more observations are needed, particularly for EHB stars in globular clusters, where our understanding of the pulsational properties is still in its infancy.

\begin{acknowledgements}
We would like to thank St\'ephane Charpinet and Marilyn Latour for many interesting discussions on the research presented here. We also gratefully acknowledge the help of the La Silla and Paranal staff with the observations. V.V.G. is an FRS-F.N.R.S. Research Associate. Support for M.C. is provided by the Ministry for the Economy, Development, and Tourism's Programa Iniciativa Cient\'{i}fica Milenio through grant IC\,120009, awarded to the Millennium Institute of Astrophysics (MAS); by Proyecto Basal PFB-06/2007; by Proyecto FONDECYT Regular \#1141141; and by CONICYT's PCI program through grant DPI20140066. T.R.M. was supported by ST/L000733.
\end{acknowledgements}

\bibliographystyle{aa}
\bibliography{omcen}

\begin{thebibliography}{55}
\expandafter\ifx\csname natexlab\endcsname\relax\def\natexlab#1{#1}\fi

\bibitem[{{Ahmad} \& {Jeffery}(2005)}]{ahmad2005}
{Ahmad}, A. \& {Jeffery}, C.~S. 2005, \aap, 437, L51

\bibitem[{{Bill{\`e}res} {et~al.}(2000){Bill{\`e}res}, {Fontaine}, {Brassard},
  {Charpinet}, {Liebert}, \& {Saffer}}]{billeres2000}
{Bill{\`e}res}, M., {Fontaine}, G., {Brassard}, P., {et~al.} 2000, \apj, 530,
  441

\bibitem[{{Bill{\`e}res} {et~al.}(2002){Bill{\`e}res}, {Fontaine}, {Brassard},
  \& {Liebert}}]{billeres2002}
{Bill{\`e}res}, M., {Fontaine}, G., {Brassard}, P., \& {Liebert}, J. 2002,
  \apj, 578, 515

\bibitem[{{Blanchette} {et~al.}(2008){Blanchette}, {Chayer}, {Wesemael},
  {Fontaine}, {Fontaine}, {Dupuis}, {Kruk}, \& {Green}}]{blanchette2008}
{Blanchette}, J.-P., {Chayer}, P., {Wesemael}, F., {et~al.} 2008, \apj, 678,
  1329

\bibitem[{{Bloemen} {et~al.}(2014){Bloemen}, {Hu}, {Aerts}, {Dupret},
  {{\O}stensen}, {Degroote}, {M{\"u}ller-Ringat}, \& {Rauch}}]{bloemen2014}
{Bloemen}, S., {Hu}, H., {Aerts}, C., {et~al.} 2014, \aap, 569, A123

\bibitem[{{Brassard} {et~al.}(2001){Brassard}, {Fontaine}, {Bill{\`e}res},
  {Charpinet}, {Liebert}, \& {Saffer}}]{brassard2001}
{Brassard}, P., {Fontaine}, G., {Bill{\`e}res}, M., {et~al.} 2001, \apj, 563,
  1013

\bibitem[{{Brown} {et~al.}(2013){Brown}, {Landsman}, {Randall}, {Sweigart}, \&
  {Lanz}}]{brown2013}
{Brown}, T.~M., {Landsman}, W.~B., {Randall}, S.~K., {Sweigart}, A.~V., \&
  {Lanz}, T. 2013, \apjl, 777, L22

\bibitem[{{Brown} {et~al.}(2001){Brown}, {Sweigart}, {Lanz}, {Landsman}, \&
  {Hubeny}}]{brown2001}
{Brown}, T.~M., {Sweigart}, A.~V., {Lanz}, T., {Landsman}, W.~B., \& {Hubeny},
  I. 2001, \apj, 562, 368

\bibitem[{{Castellani} {et~al.}(2007){Castellani}, {Calamida}, {Bono},
  {Stetson}, {Freyhammer}, {Degl'Innocenti}, {Moroni}, {Monelli}, {Corsi},
  {Nonino}, {Buonanno}, {Caputo}, {Castellani}, {Dall'Ora}, {Del Principe},
  {Ferraro}, {Iannicola}, {Piersimoni}, {Pulone}, \& {Vuerli}}]{castellani2007}
{Castellani}, V., {Calamida}, A., {Bono}, G., {et~al.} 2007, \apj, 663, 1021

\bibitem[{{Charpinet} {et~al.}(2001){Charpinet}, {Fontaine}, \&
  {Brassard}}]{charp2001}
{Charpinet}, S., {Fontaine}, G., \& {Brassard}, P. 2001, \pasp, 113, 775

\bibitem[{{Charpinet} {et~al.}(1997){Charpinet}, {Fontaine}, {Brassard},
  {Chayer}, {Rogers}, {Iglesias}, \& {Dorman}}]{charp1997}
{Charpinet}, S., {Fontaine}, G., {Brassard}, P., {et~al.} 1997, \apjl, 483,
  L123+

\bibitem[{{Charpinet} {et~al.}(1996){Charpinet}, {Fontaine}, {Brassard}, \&
  {Dorman}}]{charp1996}
{Charpinet}, S., {Fontaine}, G., {Brassard}, P., \& {Dorman}, B. 1996, \apjl,
  471, L103+

\bibitem[{{Charpinet} {et~al.}(2015){Charpinet}, {Van Grootel}, {Fontaine},
  {Brassard}, {Randall}, \& {Green}}]{charp2015}
{Charpinet}, S., {Van Grootel}, V., {Fontaine}, G., {et~al.} 2015, IAU General
  Assembly, 22, 54919

\bibitem[{{Charpinet} {et~al.}(2011){Charpinet}, {van Grootel}, {Fontaine},
  {Green}, {Brassard}, {Randall}, {Silvotti}, {{\O}stensen}, {Kjeldsen},
  {Christensen-Dalsgaard}, {Kawaler}, {Clarke}, {Li}, \& {Wohler}}]{charp2011}
{Charpinet}, S., {van Grootel}, V., {Fontaine}, G., {et~al.} 2011, \aap, 530,
  A3+

\bibitem[{{Clausen} \& {Wade}(2011)}]{clausen2011}
{Clausen}, D. \& {Wade}, R.~A. 2011, \apjl, 733, L42

\bibitem[{{D'Antona} {et~al.}(2002){D'Antona}, {Caloi}, {Montalb{\'a}n},
  {Ventura}, \& {Gratton}}]{d'antona2002}
{D'Antona}, F., {Caloi}, V., {Montalb{\'a}n}, J., {Ventura}, P., \& {Gratton},
  R. 2002, \aap, 395, 69

\bibitem[{{D'Cruz} {et~al.}(1996){D'Cruz}, {Dorman}, {Rood}, \&
  {O'Connell}}]{d'cruz1996}
{D'Cruz}, N.~L., {Dorman}, B., {Rood}, R.~T., \& {O'Connell}, R.~W. 1996, \apj,
  466, 359

\bibitem[{{Dhillon} {et~al.}(2007){Dhillon}, {Marsh}, {Stevenson}, {Atkinson},
  {Kerry}, {Peacocke}, {Vick}, {Beard}, {Ives}, {Lunney}, {McLay}, {Tierney},
  {Kelly}, {Littlefair}, {Nicholson}, {Pashley}, {Harlaftis}, \&
  {O'Brien}}]{dhillon2007}
{Dhillon}, V.~S., {Marsh}, T.~R., {Stevenson}, M.~J., {et~al.} 2007, \mnras,
  378, 825

\bibitem[{{Dorman} {et~al.}(1993){Dorman}, {Rood}, \& {O'Connell}}]{dorman1993}
{Dorman}, B., {Rood}, R.~T., \& {O'Connell}, R.~W. 1993, \apj, 419, 596

\bibitem[{{Fontaine} {et~al.}(2006){Fontaine}, {Brassard}, {Charpinet}, \&
  {Chayer}}]{fontaine2006}
{Fontaine}, G., {Brassard}, P., {Charpinet}, S., \& {Chayer}, P. 2006, \memsai,
  77, 49

\bibitem[{{Fontaine} {et~al.}(2003){Fontaine}, {Brassard}, {Charpinet},
  {Green}, {Chayer}, {Bill{\`e}res}, \& {Randall}}]{fontaine2003}
{Fontaine}, G., {Brassard}, P., {Charpinet}, S., {et~al.} 2003, \apj, 597, 518

\bibitem[{{Fontaine} {et~al.}(2012){Fontaine}, {Brassard}, {Charpinet},
  {Green}, {Randall}, \& {Van Grootel}}]{fontaine2012}
{Fontaine}, G., {Brassard}, P., {Charpinet}, S., {et~al.} 2012, \aap, 539, A12

\bibitem[{{Fontaine} {et~al.}(2008){Fontaine}, {Brassard}, {Green}, {Chayer},
  {Charpinet}, {Andersen}, \& {Portouw}}]{fontaine2008}
{Fontaine}, G., {Brassard}, P., {Green}, E.~M., {et~al.} 2008, \aap, 486, L39

\bibitem[{{Geier} {et~al.}(2013){Geier}, {Marsh}, {Wang}, {Dunlap}, {Barlow},
  {Schaffenroth}, {Chen}, {Irrgang}, {Maxted}, {Ziegerer}, {Kupfer},
  {Miszalski}, {Heber}, {Han}, {Shporer}, {Telting}, {G{\"a}nsicke},
  {{\O}stensen}, {O'Toole}, \& {Napiwotzki}}]{geier2013}
{Geier}, S., {Marsh}, T.~R., {Wang}, B., {et~al.} 2013, \aap, 554, A54

\bibitem[{{Green} {et~al.}(2003){Green}, {Fontaine}, {Reed}, {Callerame},
  {Seitenzahl}, {White}, {Hyde}, {{\O}stensen}, {Cordes}, {Brassard}, {Falter},
  {Jeffery}, {Dreizler}, {Schuh}, {Giovanni}, {Edelmann}, {Rigby}, \&
  {Bronowska}}]{green2003}
{Green}, E.~M., {Fontaine}, G., {Reed}, M.~D., {et~al.} 2003, \apjl, 583, L31

\bibitem[{{Han}(2008)}]{han2008}
{Han}, Z. 2008, \aap, 484, L31

\bibitem[{{Han} {et~al.}(2003){Han}, {Podsiadlowski}, {Maxted}, \&
  {Marsh}}]{han2003}
{Han}, Z., {Podsiadlowski}, P., {Maxted}, P.~F.~L., \& {Marsh}, T.~R. 2003,
  \mnras, 341, 669

\bibitem[{{Han} {et~al.}(2002){Han}, {Podsiadlowski}, {Maxted}, {Marsh}, \&
  {Ivanova}}]{han2002}
{Han}, Z., {Podsiadlowski}, P., {Maxted}, P.~F.~L., {Marsh}, T.~R., \&
  {Ivanova}, N. 2002, \mnras, 336, 449

\bibitem[{{Hu} {et~al.}(2011){Hu}, {Tout}, {Glebbeek}, \& {Dupret}}]{hu2011}
{Hu}, H., {Tout}, C.~A., {Glebbeek}, E., \& {Dupret}, M.-A. 2011, \mnras, 418,
  195

\bibitem[{{Jeffery} \& {Saio}(2006)}]{jeffery2006}
{Jeffery}, C.~S. \& {Saio}, H. 2006, \mnras, 372, L48

\bibitem[{{Johnson} {et~al.}(2014){Johnson}, {Green}, {Wallace}, {O'Malley},
  {Amaya}, {Biddle}, \& {Fontaine}}]{johnson2014}
{Johnson}, C., {Green}, E., {Wallace}, S., {et~al.} 2014, in Astronomical
  Society of the Pacific Conference Series, Vol. 481, 6th Meeting on Hot
  Subdwarf Stars and Related Objects, ed. V.~{van Grootel}, E.~{Green},
  G.~{Fontaine}, \& S.~{Charpinet}, 153

\bibitem[{{Justham} {et~al.}(2011){Justham}, {Podsiadlowski}, \&
  {Han}}]{justham2011}
{Justham}, S., {Podsiadlowski}, P., \& {Han}, Z. 2011, \mnras, 410, 984

\bibitem[{{Kilkenny}(2010)}]{kilkenny2010}
{Kilkenny}, D. 2010, \apss, 329, 175

\bibitem[{{Kilkenny} {et~al.}(1997){Kilkenny}, {Koen}, {O'Donoghue}, \&
  {Stobie}}]{kilkenny1997}
{Kilkenny}, D., {Koen}, C., {O'Donoghue}, D., \& {Stobie}, R.~S. 1997, \mnras,
  285, 640

\bibitem[{{Kupfer} {et~al.}(2015){Kupfer}, {Geier}, {Heber}, {{\O}stensen},
  {Barlow}, {Maxted}, {Heuser}, {Schaffenroth}, \& {G{\"a}nsicke}}]{kupfer2015}
{Kupfer}, T., {Geier}, S., {Heber}, U., {et~al.} 2015, \aap, 576, A44

\bibitem[{{Latour} {et~al.}(2011){Latour}, {Fontaine}, {Brassard}, {Green},
  {Chayer}, \& {Randall}}]{latour2011}
{Latour}, M., {Fontaine}, G., {Brassard}, P., {et~al.} 2011, \apj, 733, 100

\bibitem[{{Latour} {et~al.}(2015){Latour}, {Fontaine}, {Green}, \&
  {Brassard}}]{latour2015}
{Latour}, M., {Fontaine}, G., {Green}, E.~M., \& {Brassard}, P. 2015, \aap,
  579, A39

\bibitem[{{Latour} {et~al.}(2014){Latour}, {Randall}, {Fontaine}, {Bono},
  {Calamida}, \& {Brassard}}]{latour2014}
{Latour}, M., {Randall}, S.~K., {Fontaine}, G., {et~al.} 2014, \apj, 795, 106

\bibitem[{{Maxted} {et~al.}(2001){Maxted}, {Heber}, {Marsh}, \&
  {North}}]{maxted2001}
{Maxted}, P.~f.~L., {Heber}, U., {Marsh}, T.~R., \& {North}, R.~C. 2001,
  \mnras, 326, 1391

\bibitem[{{Miller Bertolami} {et~al.}(2008){Miller Bertolami}, {Althaus},
  {Unglaub}, \& {Weiss}}]{miller2008}
{Miller Bertolami}, M.~M., {Althaus}, L.~G., {Unglaub}, K., \& {Weiss}, A.
  2008, \aap, 491, 253

\bibitem[{{Moehler} {et~al.}(2011){Moehler}, {Dreizler}, {Lanz}, {Bono},
  {Sweigart}, {Calamida}, \& {Nonino}}]{moehler2011}
{Moehler}, S., {Dreizler}, S., {Lanz}, T., {et~al.} 2011, \aap, 526, A136+

\bibitem[{{Moni Bidin} {et~al.}(2008){Moni Bidin}, {Catelan}, \&
  {Altmann}}]{monibidin2008}
{Moni Bidin}, C., {Catelan}, M., \& {Altmann}, M. 2008, \aap, 480, L1

\bibitem[{{Moni Bidin} {et~al.}(2015){Moni Bidin}, {Momany}, {Montalto},
  {Catelan}, {Villanova}, {Piotto}, \& {Geisler}}]{monibidin2015}
{Moni Bidin}, C., {Momany}, Y., {Montalto}, M., {et~al.} 2015, \apjl, 812, L31

\bibitem[{{Napiwotzki}(1993)}]{napi1993}
{Napiwotzki}, R. 1993, \actaa, 43, 343

\bibitem[{{{\O}stensen} {et~al.}(2010){{\O}stensen}, {Oreiro}, {Solheim},
  {Heber}, {Silvotti}, {Gonz{\'a}lez-P{\'e}rez}, {Ulla}, {P{\'e}rez
  Hern{\'a}ndez}, {Rodr{\'{\i}}guez-L{\'o}pez}, \& {Telting}}]{ostensen2010}
{{\O}stensen}, R.~H., {Oreiro}, R., {Solheim}, J.-E., {et~al.} 2010, \aap, 513,
  A6+

\bibitem[{{Randall} {et~al.}(2015){Randall}, {Bagnulo}, {Ziegerer}, {Geier}, \&
  {Fontaine}}]{randall2015}
{Randall}, S.~K., {Bagnulo}, S., {Ziegerer}, E., {Geier}, S., \& {Fontaine}, G.
  2015, \aap, 576, A65

\bibitem[{{Randall} {et~al.}(2009){Randall}, {Calamida}, \&
  {Bono}}]{randall2009}
{Randall}, S.~K., {Calamida}, A., \& {Bono}, G. 2009, \aap, 494, 1053

\bibitem[{{Randall} {et~al.}(2011){Randall}, {Calamida}, {Fontaine}, {Bono}, \&
  {Brassard}}]{randall2011}
{Randall}, S.~K., {Calamida}, A., {Fontaine}, G., {Bono}, G., \& {Brassard}, P.
  2011, \apjl, 737, L27

\bibitem[{{Schaffenroth} {et~al.}(2014){Schaffenroth}, {Geier}, {Heber},
  {Kupfer}, {Ziegerer}, {Heuser}, {Classen}, \& {Cordes}}]{schaffenroth2014}
{Schaffenroth}, V., {Geier}, S., {Heber}, U., {et~al.} 2014, \aap, 564, A98

\bibitem[{{Schuh} {et~al.}(2006){Schuh}, {Huber}, {Dreizler}, {Heber},
  {O'Toole}, {Green}, \& {Fontaine}}]{schuh2006}
{Schuh}, S., {Huber}, J., {Dreizler}, S., {et~al.} 2006, \aap, 445, L31

\bibitem[{{Stetson}(1987)}]{stetson1987}
{Stetson}, P.~B. 1987, \pasp, 99, 191

\bibitem[{{Stetson}(1994)}]{stetson1994}
{Stetson}, P.~B. 1994, \pasp, 106, 250

\bibitem[{{Van Grootel} {et~al.}(2010{\natexlab{a}}){Van Grootel}, {Charpinet},
  {Fontaine}, {Brassard}, {Green}, {Randall}, {Silvotti}, {{\O}stensen},
  {Kjeldsen}, {Christensen-Dalsgaard}, {Borucki}, \& {Koch}}]{val2010a}
{Van Grootel}, V., {Charpinet}, S., {Fontaine}, G., {et~al.}
  2010{\natexlab{a}}, \apjl, 718, L97

\bibitem[{{Van Grootel} {et~al.}(2010{\natexlab{b}}){Van Grootel}, {Charpinet},
  {Fontaine}, {Green}, \& {Brassard}}]{val2010b}
{Van Grootel}, V., {Charpinet}, S., {Fontaine}, G., {Green}, E.~M., \&
  {Brassard}, P. 2010{\natexlab{b}}, \aap, 524, A63

\bibitem[{{Woudt} {et~al.}(2006){Woudt}, {Kilkenny}, {Zietsman}, {Warner},
  {Loaring}, {Copley}, {Kniazev}, {V{\"a}is{\"a}nen}, {Still}, {Stobie},
  {Burgh}, {Nordsieck}, {Percival}, {O'Donoghue}, \& {Buckley}}]{woudt2006}
{Woudt}, P.~A., {Kilkenny}, D., {Zietsman}, E., {et~al.} 2006, \mnras, 371,
  1497

\end{thebibliography}

\end{document}